\documentclass[preprint,12pt,3p,authoryear]{elsarticle}

\setcitestyle{authoryear,open={(},close={)}}

\usepackage{amssymb}
\usepackage{subfig}
\usepackage{color}
\usepackage{hyperref}
\usepackage{lineno}
\usepackage[title]{appendix}  

\usepackage{natbib}

\begin{document}

\begin{frontmatter}

\title{Herding boosts too-connected-to-fail risk in stock market of China}

\author[label1]{Shan Lu}
\author[label1]{Jichang Zhao\corref{cor1}}
\ead{jichang@buaa.edu.cn}
\author[label1]{Huiwen Wang} 
\author[label1]{Ruoen Ren} 

\address[label2]{School of Economics and Management, Beihang University, Beijing, China}
\cortext[cor1]{Corresponding author}

\begin{abstract}
The crowd panic and its contagion play non-negligible roles at the time of the stock crash, especially for China where inexperienced investors dominate the market. However, existing models rarely consider investors in networking stocks and accordingly miss the exact knowledge of how panic contagion leads to abrupt crash. In this paper, by networking stocks of sharing common mutual funds, a new methodology of investigating the market crash is presented. It is surprisingly revealed that the herding, which origins in the mimic of seeking for high diversity across investment strategies to lower individual risk, will produce too-connected-to-fail stocks and reluctantly boosts the systemic risk of the entire market. Though too-connected stocks might be relatively stable during the crisis, they are so influential that a small downward fluctuation will cascade to trigger severe drops of massive successor stocks, implying that their falls might be unexpectedly amplified by the collective panic and result in the market crash. Our findings suggest that the whole picture of portfolio strategy has to be carefully supervised to reshape the stock network.

\end{abstract}


\end{frontmatter}

\section{Introduction}
The stock market of China experienced a boom-and-bust in 2015. From the beginning of 2015, good news about the stock market as China economic vehicles published on mainstream media, attracting more than 30 million influx of new individual investors into the market, most of which were inexperienced~\citep{sun2017predicting}, yet deeply believed that the market would keep going up, inducing a trend of impulsive buying and severe overvaluation in the bull market~\citep{zhou2017}. The turbulence started with the popping of the stock market bubble on June 12 2015, when Shanghai Stock Exchange Composite (SSEC) Index reached over the 5000 points for the first time in past eight years. Thereafter, the SSEC Index began its abrupt downward spiral, with a major systemic aftershock occurred in 26 June, when over 2000 stocks listed on the two exchanges hit their price limits (in China stock market, the allowed maximum one-day drop of a stock is ten percent of its closing price last day). Those limits prevented a number of stocks from sharper declines, but they also made it nearly impossible for investors to exit positions. 

As the widespread down limits made the whole market badly short of liquidity, rumors and negative news propagated extremely fast, especially via dense social networks of investors that previously formed to share information about the bull market. People began to collectively lose faith and the confidence crushed. With little awareness of systemic risk and lacking of the relevant theoretical backups, none administrative measures had predicted such a great range of stocks `failures'. After huge amount of panic sell-off, a-third of A-shares market value was lost within one month of the catastrophic crash. The severe systemic fluctuation had smashed market confidence built by government that was supposed to make financial market as intermediation to deploy excess giving back into the real economy.

The stock market crisis experienced not only in China but also over the world has raised pressing concerns on the complex systemic risk. Nevertheless, the existing studies on stock market crash mainly give priority to time-series analysis and study the stocks movement separately~\citep{choudhry1996, amihud2009}, failing to explain the failure contagion prompted by stock inter-linkages. Many efforts have also been devoted to modeling the stock market with the language of network science~\citep{mantegna1999introduction, mantegna1999,billio2012, diebold2014, kenett2015, li2015, xu2017, wang2017, sun2017detecting, zhao2017stock}, in which the topological characteristics of the market and even its evolving dynamics have been investigated at length. But still, few of them pay sufficient attention to market systemic risk under extreme volatility. 

Meanwhile, recent studies in banking systems take rather different account of systemic risk. Instead of focusing on the individual institution intending to minimize risk and maximize profit, more attention was paid to the potential effects on the system-wide stability~\citep{schweitzer2009, haldane2011, beale2011, poledna2015}. It has been found that the individual's homogeneous strategy in reality and the heterogeneity interplay in network do have effects on systemic risk~\citep{may2008, may2010}. However, when extending these studies into stock market analysis, one essential source of systemic risk would be missed, namely the market participants' psychological factors, especially in a less mature market like China~\citep{zhou2017}.

Moreover, from the view of human nature in market crash, Shiller's questionnaires study offered survey evidence that investor fashions as a factor of crisis may well predict the market volatility~\citep{shiller1987, shiller1988,shiller1991}. But how the investor psychology and behavior possibly influence the vulnerability of market system is not well explored. In particular, one of the potential factors for market anomalies called `herding effect'~\citep{scharfstein1990,falkenstein1996,sias2004}, referring to the investors mimicking each other's strategy or following the fashion, has not been given enough attention when talking about stock market crisis.

To fill these gaps, we take connections between investors and stocks into consideration when developing the stock network to study the China market crash in June 2015. This could relate market panic to market crash, and make it possible to unfold stock market systemic failures from the view of contagious psychological phenomena and relevant behaviors. Through networking stocks with common mutual fund institutions, we find that, despite the fact that institutions diversify their portfolios to make individual safer, heterogeneity between stocks with different properties leads to institutional herding as they make similar equity holding choices on the whole. This makes some stocks highly connect to other stocks and become `super-spreaders', increasing the too-connected-to-fail risk~\citep{chan-lau2010, leon2011}. In circumstance of immature market, rumors regarding the topic easily spook investors and the market panic is promptly broadcasted through those too-connected stocks. Empirical evidence from information networks like Weibo, a Twitter's variant in China, has also demonstrated that messages with emotion of fear occupied more than 40\% of the stock-related tweets after the crisis began, especially for the individual inexperienced investors~\citep{zhou2017}. In this circumstances, those highly interconnected stocks might be strong when facing market turbulence, however, will probably lead to knock-on-effects or even a huge turnover in the system, bringing about instabilities of market~\citep{haldane2011,battiston2016, marco2017}. 

Our work would be insightful for financial market policies that favor stock market with structures more robust to economic shocks~\citep{haldane2011}. Though the herding investment strategy could benefit individual institution, regulations should pay more attention to systemic risk. One way to achieve it is to encourage heterogeneities of investors strategy to reshape the stock network, thus avoiding the too-connected-to-fail problem. 

The rest of the paper is organized as follows. Section 2 illustrates the network modeling approach that quantifying the connections among stocks based on an investor-stock bipartite graph in the first place, and then moves to data as well as data preprocessing. In section 3, we present main results. We start with the properties of the network being analyzed in the first subsection and find the disassortativeness of it. The following subsection reveals it is the herding behavior that leads to the disassortativeness, whereas creating `too-connected' stocks. Using the stock prices data during real market crash in 2015, the third subsection explores the performance of stocks with different out-degrees and finds that those too-connected stocks are more stable in the crash. In addition, when the too-connected stocks dropped a little, its successors in the network declined more than that. The Granger Causality tests carried out in the fourth subsection show that the drops in prices of too-connected stocks provide statistically significant information about the drops in prices of their successors. Given the results in section 3, section 4 develops broad discussion on the findings, along with the limitations and extensions of this study. Finally, we draw conclusions in section 5.

\section{Network modeling}

\subsection{Motivations}

Most existing models forge links between stocks mainly based on the similarity of their time-series~\citep{mantegna1999, billio2012, diebold2014, kenett2015, xu2017, wang2017}. While measures of connections in other economic network modelings that could reflect the actual interactions~\citep{soramaki2007,hidalgo2009,chan-lau2010, vitali2011,leon2011,battiston2012, zhang2016} have not yet been extended to the study of networking stock market. In stock market of China, the actual interactions coupled within stocks are pretty important. When exposing one stock to being sell-off to the decline limit (the allowed maximum one-day drop of a stock is ten percent of its closing price last day), the investors who hold the stock suffer from holding value decrement and being not able to draw this part of investment out of the market. They could run scared easily by selling other stocks in hand because of the need for liquidity, the anxiety of further loss, and even the general loss of confidence. In addition to panic selling, many investors who faced margin calls on their stocks were forced to sell off shares during the crash happened in 2015. These would precipitate price falling of other stocks. 

\subsection{Networking method}

\begin{figure}[htbp!]
\centering
\includegraphics[width= 0.5\linewidth]{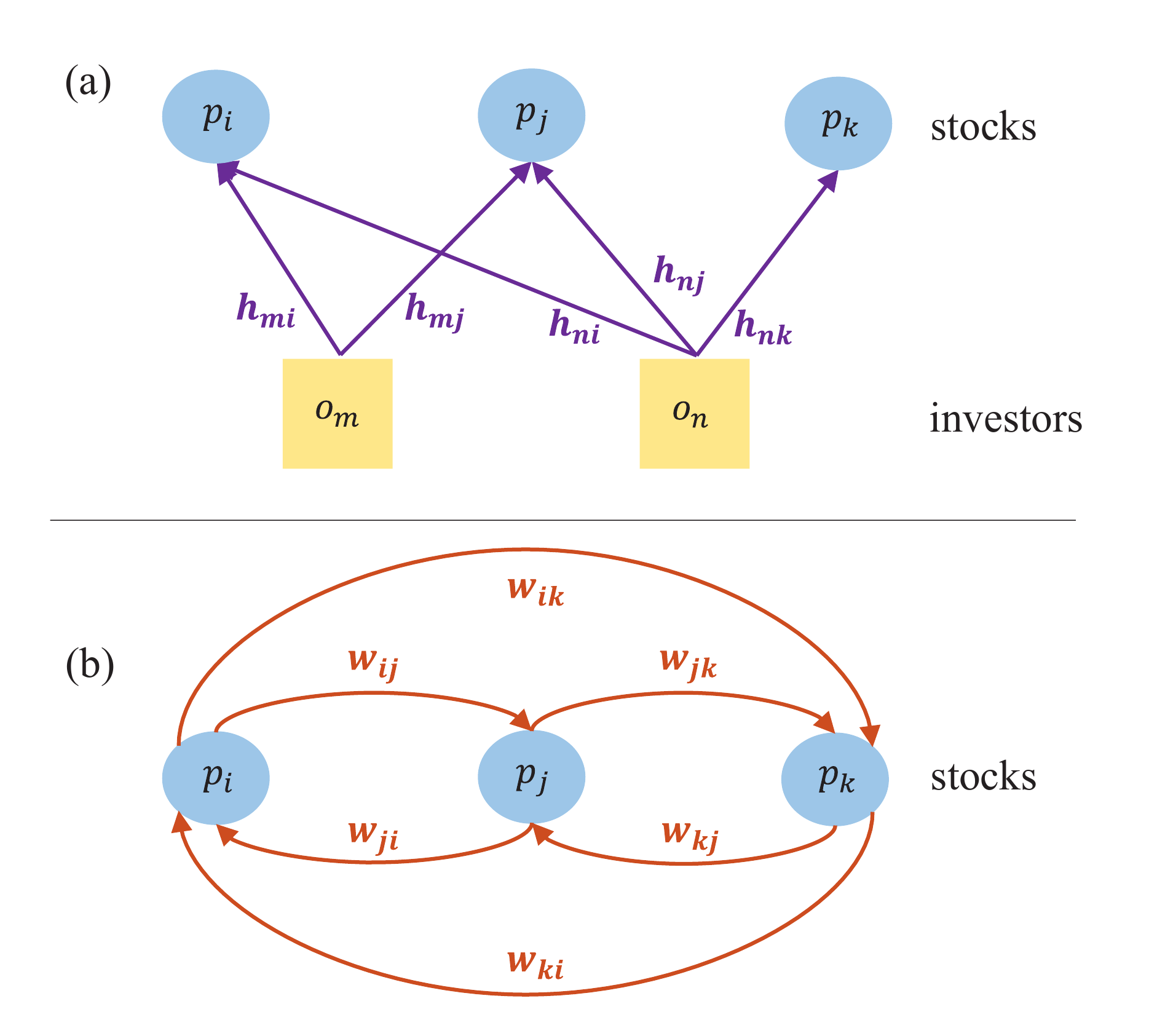}
\caption{\bf{Modeling stock network through common investors.} \normalfont{(a) is the bipartite graph between stocks and investors, in which the link weight $h_{mi}$ between investor $o_m$ and stock $p_i$ is defined as the market value of stock $p_i$ owned by investor $o_m.$ Denote $C_{ij}=$\{common investors of stock $p_i$ and $p_j$\}. (b) is the stock network projected from the bipartite graph, in which the weight $w_{ij}$ of the directed link from stock $p_i$ to $p_j$ is defined as the sum of market value of stock $p_i$ owned by all their common investors, i.e., $w_{ij}=\sum_{m\in{C_{ij}}}h_{mi},$ similarly $w_{ji}=\sum_{m\in{C_{ij}}}h_{mj}$.}}
\label{fig:projection}
\end{figure}

Motivated by the above, the idea here is that stocks connect to each other through common investors, as illustrated in Fig.~\ref{fig:projection}(a). A weighted bipartite network between investors and stocks is denoted as $B=\{O,P,E,H\}$. Here $O$ is investor vertex set. The node of investor $m$ is denoted as $o_m$, where $o_m\in{O}$ . $P$ is the vertex set of stocks. The node of stock $i$ is denoted as $p_i$, where $p_i\in{P}$. $E$ is the set of edges between $O$ and $P$. Once investor $m$ invests in stock $i$, there is a link named $e_{mi}$ established between $o_m$ and $p_i$, where $e_{mi}\in{E}$. $H$ is the set of edge weights, where weight $h_{mi}$ of link $e_{mi}$ is the amount of stock $i$'s market value owned by investor $m$. We have $h_{mi}\in{H}$.

For the convenience of directly showing the relations among stocks, the bipartite graph is compressed by projection. Denote the stocks network as $V=\{P,F,W\}$, see Fig.~\ref{fig:projection}(b). Still, $P$ is the vertex set of stocks. $F$ is the edges set and $W$ is edge weights set. The simplest way to project the bipartite network is that one stock is connected to another stock only when they share at least one common investor. Suppose that in the bipartite graph $B$, there are $n_{ij}$ investors hold both stock $i$ and $j$. Denote $C_{ij}=$\{common investors of stock $p_i$ and $p_j$\}, $n_{ij}>0$, and $n_{ij}=n_{ji}$. Then there is a directed link $f_{ij}$ from $p_i$ to $p_j$ and another directed link $f_{ji}$ from $p_j$ to $p_i$. And their weights are denotes as $w_{ij}$ and $w_{ji}$, respectively, with $f_{ij}, f_{ji}\in{F}$ and $w_{ij}, w_{ji}\in{W}$. To evaluate the influence of one node have on another, the weight $w_{ij}$ of link $f_{ij}$ is defined as the sum of market value of stock $i$ owned by investors in $C_{ij}$, i.e. $w_{ij}=\sum_{m\in{C_{ij}}}h_{mi}$. Same with stock $p_j$, i.e. $w_{ji}=\sum_{m\in{C_{ij}}}h_{mj}$. And usually we have $w_{ij}\ne{w_{ji}}$. This mean that, when common investors of stocks $i$ and $j$ put more money on stock $i$, the degree of the impact of drops in price of stock $i$ on stock $j$ is greater than the opposite direction. 

To our best knowledge, it is the first time for a stock network to be established based on investment ownership relationship. Unlike most undirected correlation-based graphs, in our model, the market is represented by a directed network with edges indicating not only the direction of impact between two stocks but also how much the influence is. More importantly, the linkages explicitly reflect the asymmetric interplay between stocks that induced by panic selling activity of investors. Therefore, the approach offers descent support for discussion on systemic risk from the prospective of selling panic contagion of investors. And our model can be deployed to any of other stock markets, either globally or nationally without modification. 

\subsection{Data}

\subsubsection{Links between stocks and investors}

We considered using mutual funds as proxy of market investors. The reason is that most new individual investors in China are unprofessional and they are easily allured by mutual fund institutions' holding position and investment trending. Besides, mutual funds in China are short term institutional investors for holding for a short period~\citep{yuan2008} and may sometimes behave more like the general public~\citep{falkenstein1996}. The mean ownership statistic indicates that the mutual funds own an average of 5.5 percent of shares in June 2015, hence the entire sample of mutual funds represents a small yet significant portion of total equity ownership of the A-shares market. All of the above provides rationality for us to use mutual fund institutions to approximate stock investors. 

The dataset of mutual funds' stock holdings in China is a snapshot on June 30 2015, around the period that the severe stock market crash happened. The dataset is provided by the Wind Information (Wind Info), a leading integrated service provider of financial data in China. Though the ownership data is taken at one particular time of the year, it represents noisy yet unbiased estimate of mutual funds' investment preferences in that year or at least days around the reporting date, because the holdings stay the same for a certain period \citep{yuan2008}. The raw dataset contains the market value of shares held by each mutual fund for listed stocks on June 30 2015 and covers 1512 mutual funds and 2709 stocks listed on either Shanghai Stock Exchange or Shenzhen Stock Exchange. The mutual funds are all those disclosed their holdings in China. As for the stocks, it should be claimed that the number of stocks listed in the A-share market are 2956 but some stocks has no investment from any mutual funds, and are excluded in the raw dataset. Overall, the dataset is of enough representative of investing pattern in the stock market of China.

Before raising a weighted bipartite graph as sketched in Fig.~\ref{fig:projection}(a), we group ownership by mutual fund management companies as we assume that mutual funds under the same management company could have collective actions on an individual stock \citep{yuan2008}. Thus, instead of raising multiple links between mutual funds managed by the same management institute and their holding stocks, we only make links between the management institutions and stocks. The edges weight are equal to the sum of market value hold by mutual funds under the same management institutes. After necessary data preprocessing, there turns out to be 87 mutual fund institutions and 2709 stocks included in the following study. What we finally have for building up the network are data revealing the market value of stock $i$ that hold by mutual fund institution $m$, which is just what we denoted as $h_{mi}$ in Fig.~\ref{fig:projection}(a).

\subsubsection{Stock prices}

The data of stock prices are downloaded from Thomson Reuters' Tick History. Two types of price time series data are used, including \emph{end-of-day} and \emph{intraday}. We use the last price of every one minute from the \emph{intrady} data in order to calculate the percentage changes of stocks. The price of stock $i$ at time $t$ on day $d$ is $p_{i,t,d}$. We use the last price of a day from the \emph{end-of-day} data. It is the baseline price for computing the return next day, denoted as $p_{i,d}$. The percentage change of stock $i$ at time $t$ on day $d$ is computed by $(p_{i,t,d}-p_{i,d-1})/p_{i,d-1}$. The reason is that this kind of percentage changes is consistent with what investors see during trading hours on any trading information board, which could stir up tensions and impact the prices of stocks through trading behavior directly. 

As we focus on studying the market crash, only four days when big shocks happened are included in sample for the main results. These are June 26, June 29, July 2 and July 3. These four days all witnessed the abrupt decline of a great range of stocks, see Table~\ref{tab:4days}. 

\begin{table}[htbp!]
\centering
\caption{\bf{Numbers of stocks that reached their down limit.}}\label{tab:4days}
\medskip
\begin{tabular}{rccc}
  \hline
date & \shortstack{number of stocks that reached \\their lower limit}  & \shortstack{Market index percentage \\change(SSEC)} \\ 
\hline
06/26/2015 & 2027 & -7.4\%\\ 
06/29/2015 & 1514 & -3.3\%\\ 
07/02/2015 & 1462 & -3.5\%\\ 
07/03/2015 & 1429 & -5.8\%\\ 
   \hline 
\end{tabular}
\end{table}

It is worth noting that other than the market crash period, we also consider a comparison case when there was a bull market to make further discussions on our findings. Those analysis could be found in appendix, whereas the data used are from the same source and under the same preprocessing procedure as stated above. The datasets analyzed in the current study are available in the figshare.com repository, see https://doi.org/10.6084/m9.figshare.5217232.v1.

\section{Results}

\subsection{Too-connected stocks in the network}

The established network of the stock market possesses 2709 nodes and 313307 edges after necessary filtering (see Appendix Fig.~\ref{fig:filter} for details). It is directed, asymmetrical, with no self-connections. The basic statistical information of the network could be found in Table~\ref{tab:info} in rows of Jun-2015. This network is highly non-random and disassortative (also see Appendix Figs.~\ref{fig:degree_cdf}, \ref{fig:bubble}), especially in terms of out-degrees, much differs from the previous studies~\citep{kenett2015, xu2017, wang2017}.

Note that the Jun-2015 network is the one used in the analysis of main text. The Dec-2014 network is a comparison case, built by the stock holding data released in mutual funds' annual reports in December 31 2014, when there was a bull market. The similarity of indicators in Table~\ref{tab:info} shows that the established network structure is persistent, independent to the performance of the market to some extent, whether there is a jump or plunge.

\begin{table}[htbp!]
\caption{\bf{Statistical information of the networks.}}\label{tab:info}
\resizebox{\textwidth}{!}{\begin{tabular}{rllll}
  \hline
year & density & number of nodes & number of edges \\ 
  \hline
Dec-2014 & 0.04 & 2418 & 232905  \\ 
Jun-2015 & 0.043 & 2709 & 313307 \\ 
 \hline
 \hline
year & in-degree assortativity & out-degree assortativity & average degree \\ 
 \hline
Dec-2014 & -0.256 & -0.493 & 96.321 \\ 
Jun-2015 & -0.177 & -0.421& 115.654 \\ 
\hline
\hline
year & mean of edge weights & standard deviation of edge weights & sum of edge weights \\ 
\hline
Dec-2014 & 307807.1 & 369487.4 &  7.17E+10 \\ 
Jun-2015 & 299764.5 & 246610.8 &  9.39E+10 \\ 
\hline
\hline
year & \shortstack{number of strongly \\connected components} & \shortstack{number of weakly \\connected components} & \shortstack{size of largest strongly \\connected components} & \shortstack{size of largest weakly \\connected components}  \\ 
 \hline
Dec-2014 & 2115 & 173 & 304 & 2246  \\ 
Jun-2015& 2320 & 203 & 390 & 2507  \\ 
\hline
\end{tabular}}

\end{table}

\begin{figure}[h]
\centering
\includegraphics[width=3in, trim={0 5cm 0 5cm},clip]{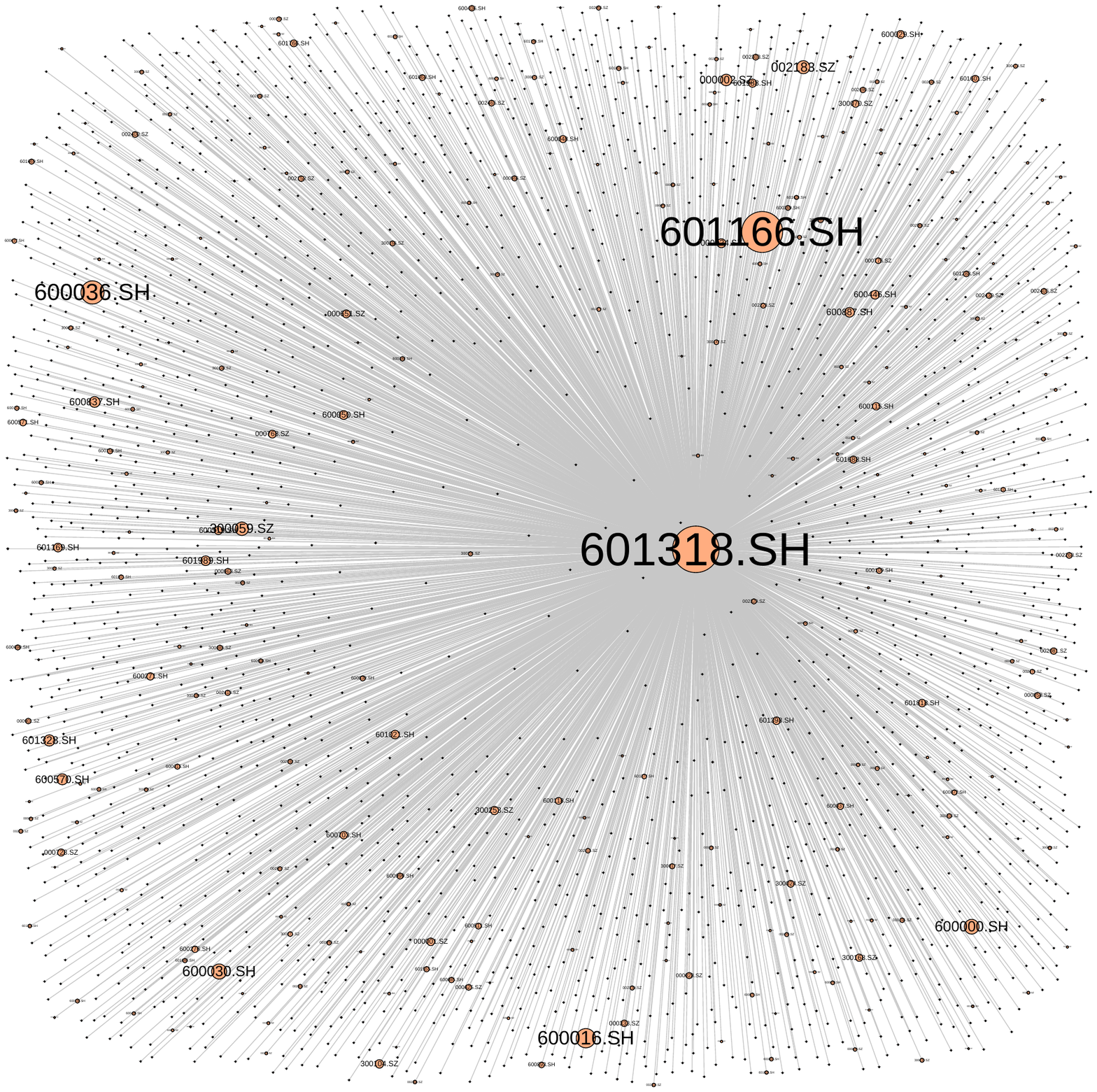}
\caption{\bf{Network visualization of a too-connected stock.} \normalfont{The size of a node is proportional to its out-degree. The subgraph contains all the edges starting from the node `601318.SH'. The ego-network illustrate how the too-connected stocks take control over the whole system.}}
\label{fig:top5}
\end{figure}

Fig.~\ref{fig:top5} shows a subgraph of the network. The subgraph contains all the edges starting from the node `601318.SH', which is the node with the most out-degrees. The graphs illustrate how the too-connected stocks take control over the whole system, raising the concern that nodes possessing many successors and their edges weight can not be ignored, tend to have great influence on other nodes with respect to both out-degrees and out-strengths in the networked market (see also Appendix Table~\ref{tab:5quantile}). Contrary to the obvious dominance in connectivity, the most connected group of stocks occupies only 18\% market value of the total 2709 securities. 

Departing from the notion of `too-big-to-fail' in economic system, the small fraction of market value but large out-degree in the network implies that it is not `big' in scale but `big' in connectivity that induces the potential issue of `too-big-to-fail' or to be more exact, `too-connected-to-fail'. Hereafter the `too-connected' stocks refer to those of high out-degrees. These stocks mainly belong to financial sector, and most of them are large-cap stocks (see Appendix Table \ref{tab:industry} and \ref{tab:style}).

Consistent with concerns in the banking system~\citep{haldane2011}, relatively small initial liquidity shocks from any of these influential nodes would make a huge contribution to systemic turnover. Recall the mechanism beyond link formation of stocks, the peculiar topology results from a specific overlap pattern among mutual funds' holdings, and provides possible paths through which the market panic spreads from one stock to its successors.

\subsection{The herding behavior of mutual funds}
\label{subsec:herding}

To fully understand where disassortativeness of the stock network origins from, we investigate mutual fund institutions' stocks holding status in this section. Traders' investment strategy may be driven by both minimizing risk for themselves and group psychology~\citep{wermers1999, sias2004}. One of the group psychology in investment decision, namely `herding', refers to that if investors have a comparative advantage in holding securities with certain characteristics, however, as a security acquires these characteristics market participants in aggregate will purchase it~\citep{falkenstein1996}. The net result is that they tend to herd together, following similar investment strategy, especially when inexperienced investors dominate the market. And this could interact with stock prices~\citep{dasgupta2011, edelen2016}. Yet, less attention has been paid to the possible effects on the stability of the entire market system. 

\begin{figure}[htbp!]
	\begin{minipage}{0.5\linewidth}
		\centering
		{\footnotesize (a)}
		\includegraphics[height = 2.8in]{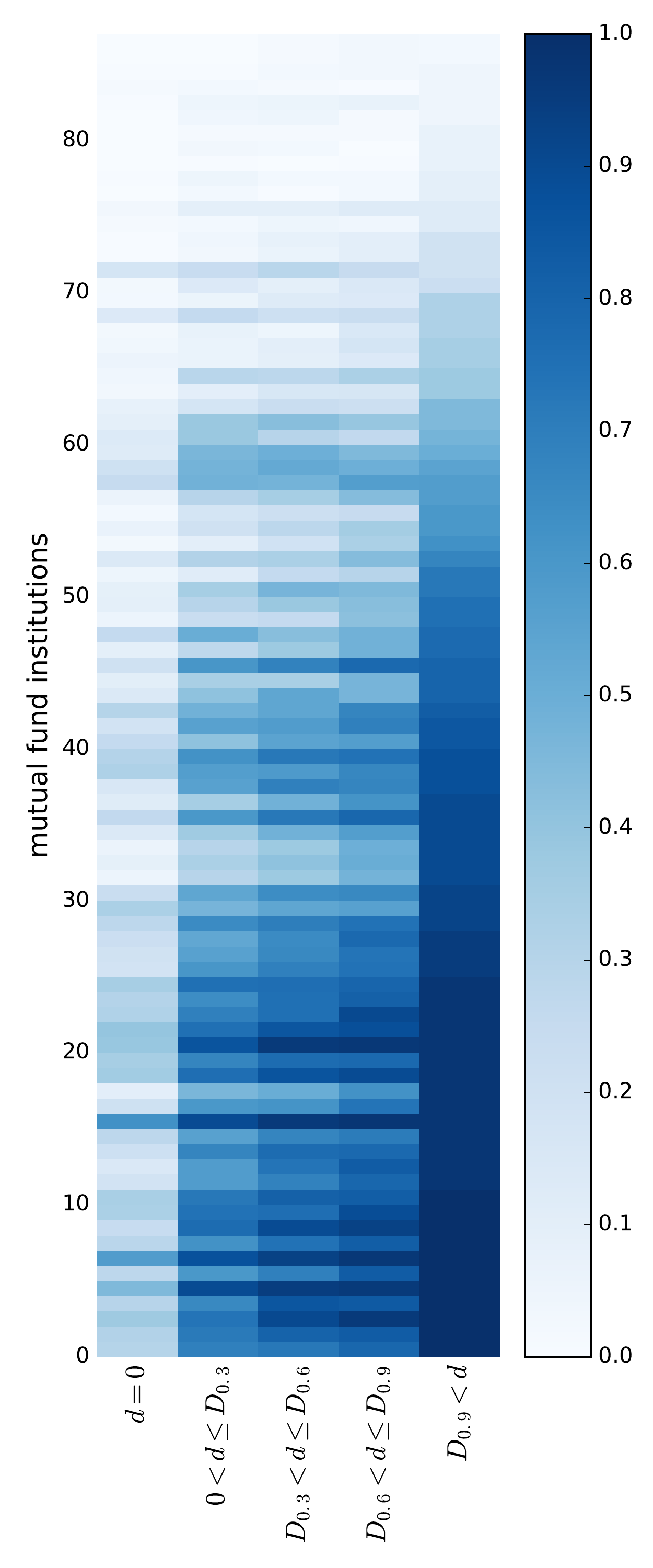}
	\end{minipage}\hfill
	\begin{minipage}{0.5\linewidth}
		\centering
		{\footnotesize (b)}
		\includegraphics[height = 2.8in]{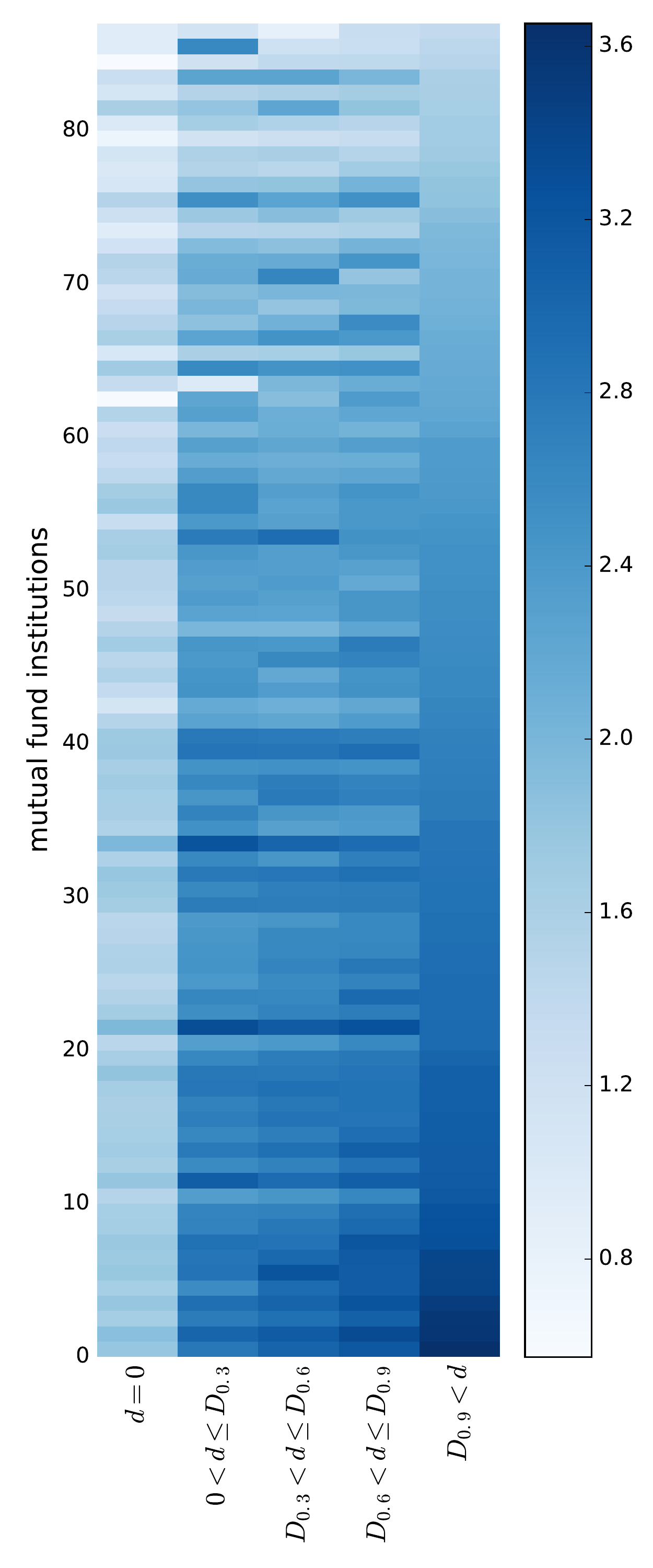}
	\end{minipage}\hfill
\caption{\bf{Mutual fund institutions' investment patterns.} \normalfont{The mutual fund institutions are labeled from 1 to 87 vertically. After excluding nodes with zero out-degrees, the 30th quantile, 60th quantile, 90th quantile of out-degree sequence are 451, 946, 1490, and denoted as $D_{0.3}$, $D_{0.6}$, $D_{0.9}$, respectively. $d$ stands for the out-degree of stocks. Every institution has one horizontal color bar showing its preference of investments. (a) Each grid presents the number of stocks one institution hold divided by the number of stocks in the corresponding out-degree category. (b) Each grid presents the average market value per stock one institution hold in the corresponding out-degree category.}}\label{fig:heat}
\end{figure}

Fig.~\ref{fig:heat} illustrates the mutual fund institutions' investment distribution in June 2015. Here, stocks are divided into five groups according to the out-degree sequence quantile, as the out-degree reflects the importance of a node in case of disturbances~\citep{chan-lau2010} and it is highly heterogeneous in the presented network. After excluding nodes with zero out-degree, the 30th quantile, 60th quantile, 90th quantile of out-degree sequence are denoted as $D_{0.3}$, $D_{0.6}$, $D_{0.9}$, respectively. $d$ stands for out-degree. The mutual fund institutions are labeled from 1 to 87 vertically in Fig.~\ref{fig:heat}. Every institution has one horizontal color bar showing its preference of investments. In Fig.~\ref{fig:heat}(a), each grid presents the number of stocks one institution hold divided by the number of stocks in the corresponding out-degree category. As can be seen in the Fig.~\ref{fig:heat}(a), horizontally, the greater the out-degree, the darker the grid, suggesting that most mutual fund institutions invest on the same stocks of highly connected. Relevant one-tailed, paired-samples $t$ tests testify that the preference increment with out-degrees is significant (see the middle column of Table~\ref{tab:t-test}). And this investment pattern exists in the stock market of China consistently without being confined by the market status (see Appendix Fig.~\ref{fig:dec2014herd}). Note that most institutions on the whole have invested a lot of stocks to diversify their investment portfolios. In particular, institutions labeled from 1 to 25 in Fig.~\ref{fig:heat}(a) invested at least 700 out of 2709 stocks. For example, China Southern Fund Management Co., Ltd alone invested 1821 stocks, ranking first in the number of stocks investing among 87 mutual fund institutions.

From a different perspective, in Fig.~\ref{fig:heat}(b), each grid presents the average market value per stock one institution hold in the corresponding out-degree category, which depicts how much each mutual fund institution invests in a certain kind of stocks on average. Similarly to Fig.~\ref{fig:heat}(a), Fig.~\ref{fig:heat}(b) shows that mutual fund institutions invest much more money per stock on stocks with larger out-degrees (relevant one-tailed, paired-samples $t$ tests are also significant, see the right column of Table~\ref{tab:t-test}), especially institutions at the bottom. Those institutions are presented in Appendix Table~\ref{tab:five_companies}, which are those invest the most money per stock in the `$D_{0.9}<d$' group, investing large amounts of money on stocks with high out-degrees while simultaneously investing many other stocks with low out-degrees. 

Both Fig.~\ref{fig:heat}(a) and Fig.~\ref{fig:heat}(b) show similar gradient color from low out-degree category to high out-degree category, implying that most mutual fund institutions have the preference on stocks with large out-degrees. Moreover, even though each mutual fund institution follows a diversified portfolio strategy whereas they have investment over all groups of stocks, surprisingly there is little diversity in mutual fund institutions' preferences as a whole. We argue that the overall similar investment pattern is a kind of herding, which origins in the mimicking investment holdings strategy. In circumstances of immature stock market in China, the institutional herding behavior influences the stock market not only through the coupling structure between institutions and stocks but also the impact of mutual funds' holding on investing choices of inexperienced investors. Though we investigate only the temporarily static snapshot of stocks holding, the results exhibit the outcomes of the trading by institutions that have been considered in previous studies~\citep{sias2004}, thus depict a meaningful aspect of herding behavior.

\begin{table}[htbp!]
\caption{\bf{One-tailed, paired-samples $t$ tests for Fig.~\ref{fig:heat}.}}\label{tab:t-test}
\medskip
\resizebox{\textwidth}{!}{\begin{tabular}{lcc}
\hline
\medskip
Hypothesis (grid value) & \shortstack{T-test Statistic\\ for Fig.~\ref{fig:heat}(a)\\($p$-value)}& \shortstack{T-test Statistic\\ for Fig.~\ref{fig:heat}(b)\\($p$-value)} \\ 
\hline
\shortstack{$H_0$:`$d=0$' no less than `$0<d\leq D_{0.3}$'€™ \\
$H_1$:`$d=0$' less than `$0<d\leq D_{0.3}$'}
				 & \shortstack{-14.48\\ (0.000*)} & \shortstack{-5.44\\(0.000*)} \\ [0.5ex]
				 \hline
\shortstack{$H_0$:`$0<d\leq D_{0.3}$' no less than `$D_{0.3}<d\leq D_{0.6}$'€™\\
			$H_1$: `$0<d\leq D_{0.3}$' less than `$D_{0.3}<d\leq D_{0.6}$'} 
				& \shortstack{-9.64\\(0.000*)} &\shortstack{-1.28\\( -0.101)} \\ [0.5ex]
				\hline
\shortstack{$H_0$: `$D_{0.3}<d\leq D_{0.6}$' no less than `$D_{0.6}<d\leq D_{0.9}$'€™\\
			$H_1$: `$D_{0.3}<d\leq D_{0.6}$' less than `$D_{0.6}<d\leq D_{0.9}$'}
				& \shortstack{-8.09\\(0.000*)} & \shortstack{-3.89\\(0.000*)}\\  [0.5ex]
				\hline
\shortstack{$H_0$: `$D_{0.6}<d\leq D_{0.9}$' no less than `$D_{0.9}<d$'€™\\
			$H_1$: `$D_{0.6}<d\leq D_{0.9}$' less than `$D_{0.9}<d$'} 
				& \shortstack{-13.12\\(0.000*)} & \shortstack{-3.742\\(0.000*)}\\ [0.5ex]
\hline
\end{tabular}}

\medskip
\scriptsize{Note: After excluding nodes with zero out-degrees, the 30th quantile, 60th quantile, 90th quantile of out-degree sequence are denoted as $D_{0.3}$, $D_{0.6}$, $D_{0.9}$, respectively. $d$ stands for out-degree. The one-tailed, paired-samples $t$ tests show the significance that the higher the out-degree, the greater the grid value is. One exception is that the `$0<d\leq D_{0.3}$' group is not significantly greater than the `$d=0$' group under the $p-value<0.05$ significance level, mainly because of the smoothing that value averaging can produce, where the `$0<d\leq D_{0.3}$' group has 117 stocks and `$d=0$' group has 2319 stocks.}
\end{table}
In particular, mutual fund institutions massively hold stocks of high out-degrees while also invest low out-degree stocks simultaneously, making it easier for some stocks to be connected to other stocks and then become `too-connected' nodes. Furthermore, the investment size on those `too-connected' stocks are large, meaning their impact on successor stocks can be fatal and hard to be ignored. Thus, those stocks, though not competent in market value, become highly connected and even too-connected-to-fail, raising the concern of domino effect in market crash.

\subsection{Too-connected stocks are more stable in the crash}
\label{subsec:hubs_stable}

The June 26 2015 experienced two thirds of stocks in A-shares market reaching their down limits, and is the second biggest crash in China stock market history as the SSEC index tumbled 345 points for its biggest one-day drop since February 2007. It is a proper sample for understanding the crash. 

\begin{figure}[htbp!]
\centering
\includegraphics[width=0.75\linewidth]{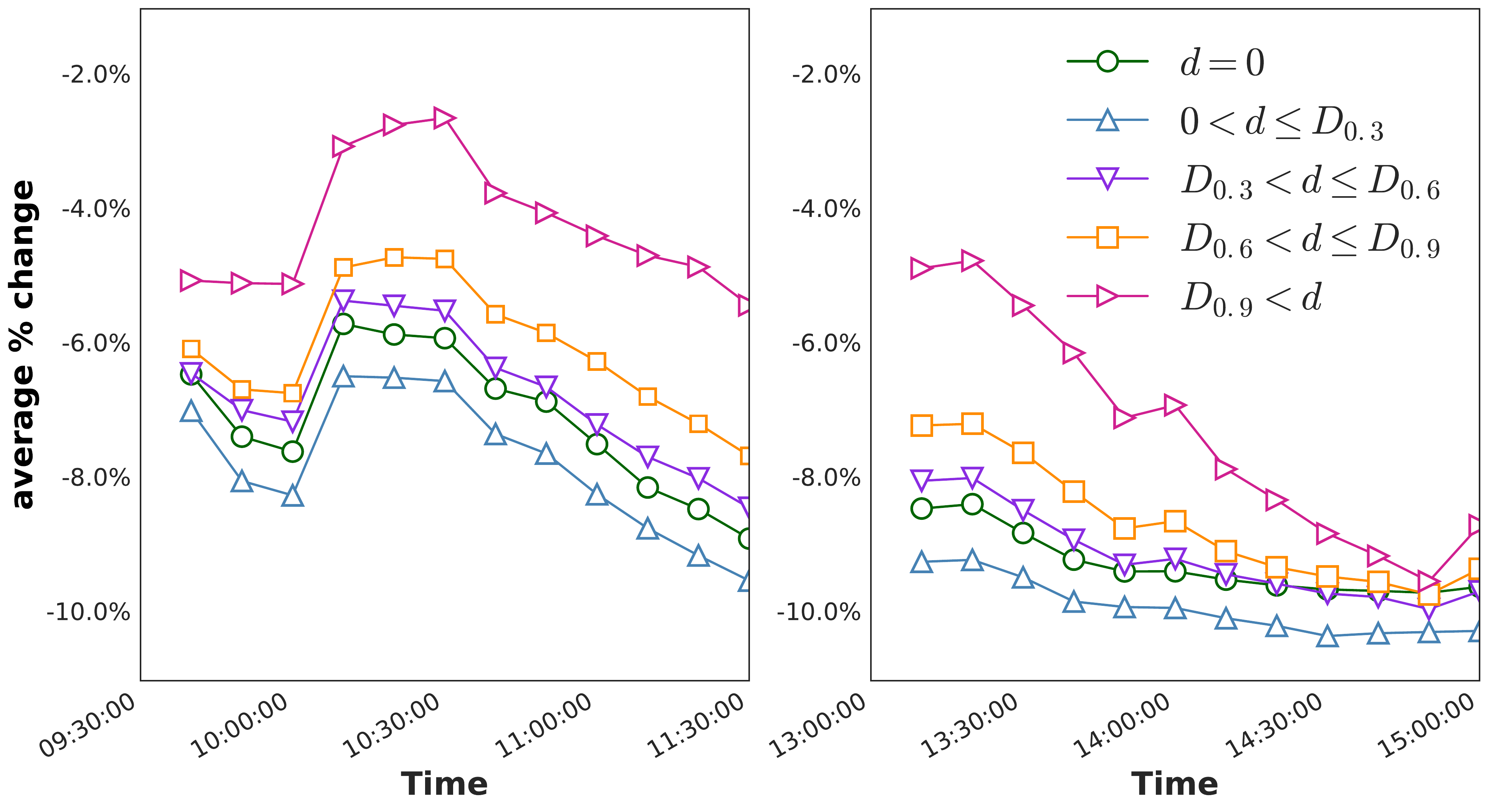}
\caption{\bf{Average percentage changes of five groups of stocks in every ten minutes on June 26, 2015.} \normalfont{Two exceptions are: 1) all groups of stocks share similar average percentage change level at the last 30 minutes of transaction hours because 2027 out of 2423 active stocks reached their lower limits when the market closed; 2) the average percentage changes of the second smallest out-degree group is lower than zero out-degree group because the latter group possesses more than 2300 stocks where the big falls and small falls are averaged. Apart from all these, Fig.~\ref{fig:626_time} generally demonstrates the greater the out-degree, the less the stock may drop when crash happens, suggesting that too-connected stocks will be relatively stable in the crash.} }\label{fig:626_time}
\end{figure}

Like what we do in Fig.~\ref{fig:heat}, we classify stocks into five groups according to their out-degrees. Fig.~\ref{fig:626_time} shows average percentage changes of the five groups of stocks on June 26 2015. To make the figure explicit to read, here we use non-overlapping time windows of ten minutes instead of one minute as described in data section, while this would neither influence computation results on these specific time stamps nor the overall trending shown in Fig.~\ref{fig:626_time}. We find that the greater the out-degree, the less the stock may drop when crash happens, suggesting that too-connected stocks will be relatively stable in the crash, especially stocks in the category of highest out-degrees (pink line) always have smaller absolute declines during the day.   

\begin{figure}[htbp!]
\centering
\includegraphics[width=.4\linewidth]{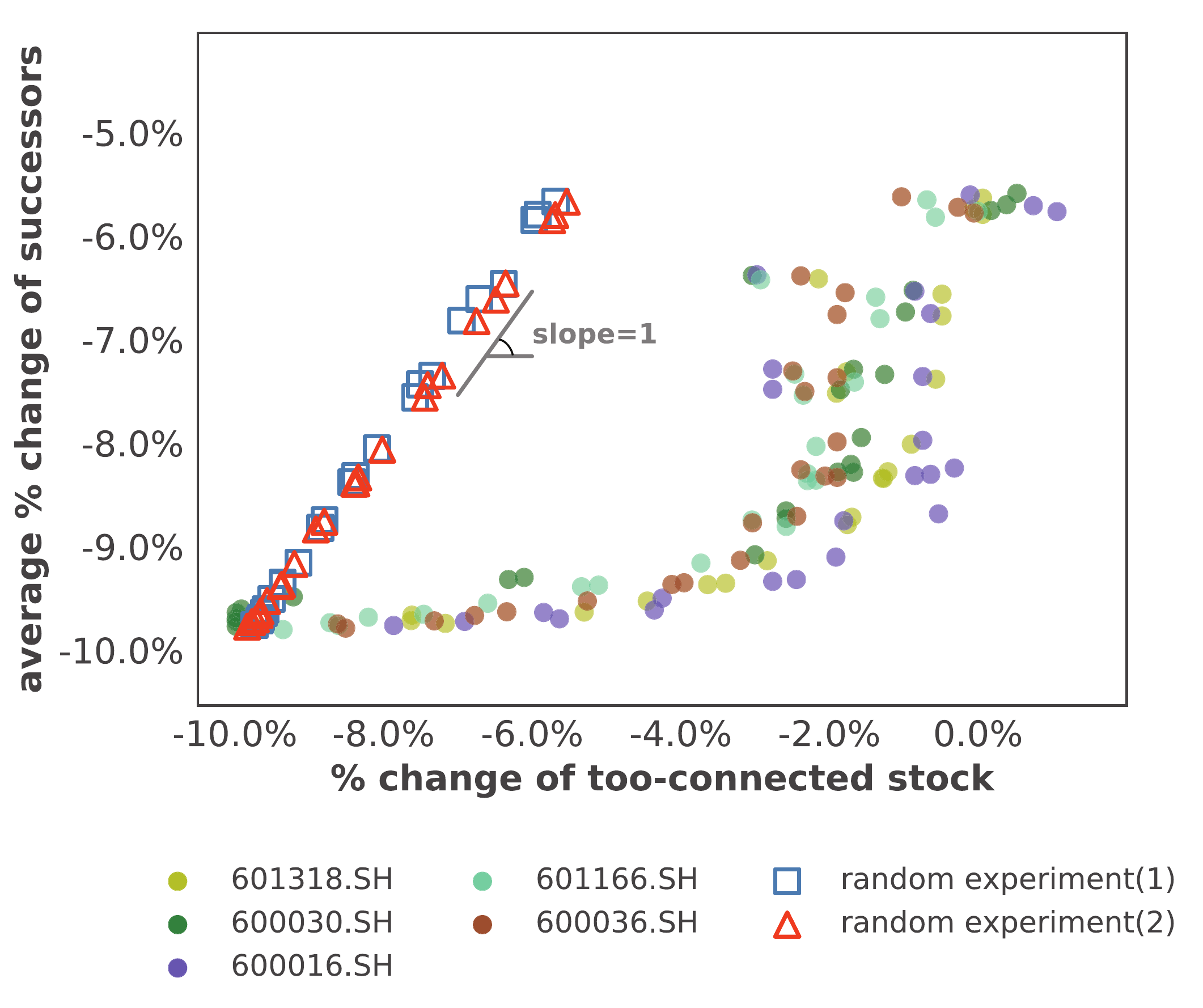}
\caption{\bf{Correlations between `too-connected' stocks' percentage changes and their successors'.} \normalfont{The average percentage changes of a stock in every ten minutes on June 26 2015 is used as its market performance, while each highly connected stock has a single percentage changes time series that consists of 24 points and corresponding successors are specified by an averaged percentage changes. Dots refer to stocks with highest out-degrees in our network. They all have more than 1900 successors. In experiment (1), a random network with 2709 nodes and 313,307 edges is built, whose degree distribution is more uniform than our network. In experiment (2), 2709 nodes of our network are randomly shuffled. In both experiments, we selected out the top five stocks of largest out-degrees in each randomized network and calculated its percentage changes as well as average percentage changes of their successors, making it comparable with dots result originated from the real network. The two random experiments are separately repeated 100 times.}}\label{fig:top5_successor}
\end{figure}

The correlation of price trend between highly connected stocks and their successors in the network is of spurt of interest. To achieve this, we investigate the percentage changes of top five stocks of largest out-degrees and their successors in Fig.~\ref{fig:top5_successor} on June 26 2015, respectively. Each dot represents the percentage change of one too-connected stock on x-axis and the average percentage change of its successors on y-axis, all in non-overlapping time windows of ten minutes. Dots in different colors tell which too-connected stock they belong to. We find that the dots are arranged in a quadratic function shape, meaning that when the prices of too-connected stocks fall a little, their successors' prices have dropped a lot.  For example, when the too-connected stocks percentage changes are -2\%, their successors have 9\% off in prices.

For the purpose of significance test, two different random experiments are performed and demonstrated in Fig.~\ref{fig:top5_successor}. In experiment (1), all links in the established stock network are randomly shuffled to reorganized the structure while keeping the density static and as compared to the original network, top stocks and their successors will be completely randomized. In experiment (2), only nodes of the established stock network are randomly shuffled to ignore the sharing of investors while keeping the network structure static, and as compared to the original network, the degree distribution will be the same, however, top stocks and their successors will be randomized. In both experiments, we selected out the top five stocks of largest out-degrees and calculated their percentage changes as well as average percentage changes of their successors, making it comparable with dots originated from the real network. The two random experiments are separately repeated 100 times and mean results of them are shown in square and triangle makers in Fig.~\ref{fig:top5_successor}.

Fig.~\ref{fig:top5_successor} exhibits that, either in random experiment (1) or (2), the percentage changes of the `too-connected' stocks are nearly the same with their successors'. This implies that under either a complete alternative network structure or reorganizations of nodes only, which could reflect distinct landscapes of investment strategies, the highly connected nodes' percentage changes behave obviously different with our empirical outcomes. Thus the empirical results significantly demonstrate that the most connected nodes fell less than their successors in most transaction hours. In other words, when the too-connected stocks dropped a little, their successors declined more than that, suggesting that the turbulence of the market might be amplified by the too-connected stocks and this phenomena is pervasive and can be well replicated in other three crisis days (see Appendix Fig.~\ref{fig:crisis3days_return},\ref{fig:crisis3days_scatter}).

\subsection{Too-connected stocks boost the market risk}
\label{subsec:boost_risk}

Whether the too-connected stocks are useful in forecasting their successors is remained to be discussed. To answer this, time series data of stocks' percentage changes in every one minute are utilized to testify the influence mechanism through Granger Causality test. Granger Causality test tells the prediction ability of a time series $X$ on another time series $Y$. $X$ is Granger-cause $Y$ if $Y$ could be better predicted by the historical values of both $X$ and $Y$ than only by historical values of $Y$ \citep{granger1969investigating}. We follow the procedure of~\citep{toda1995statistical} for Granger Causality testing because of its advantage of avoiding a pretest bias. It could be carried out through building prediction models and Wald-tests. For each one of the too-connected stocks, denote its intraday-one-minute percentage changes as $X$ and one of its successors' as $Y$. The null hypothesis here is that $X$ is not Granger-cause $Y$. We carry out Granger Causality tests on all pairs of stocks in a given day. The significance level $p$ of tests is set to 0.05. If $p<0.05$, the null hypothesis is rejected. 

\begin{figure}[htbp!]
\centering
\includegraphics[width=0.5\linewidth]{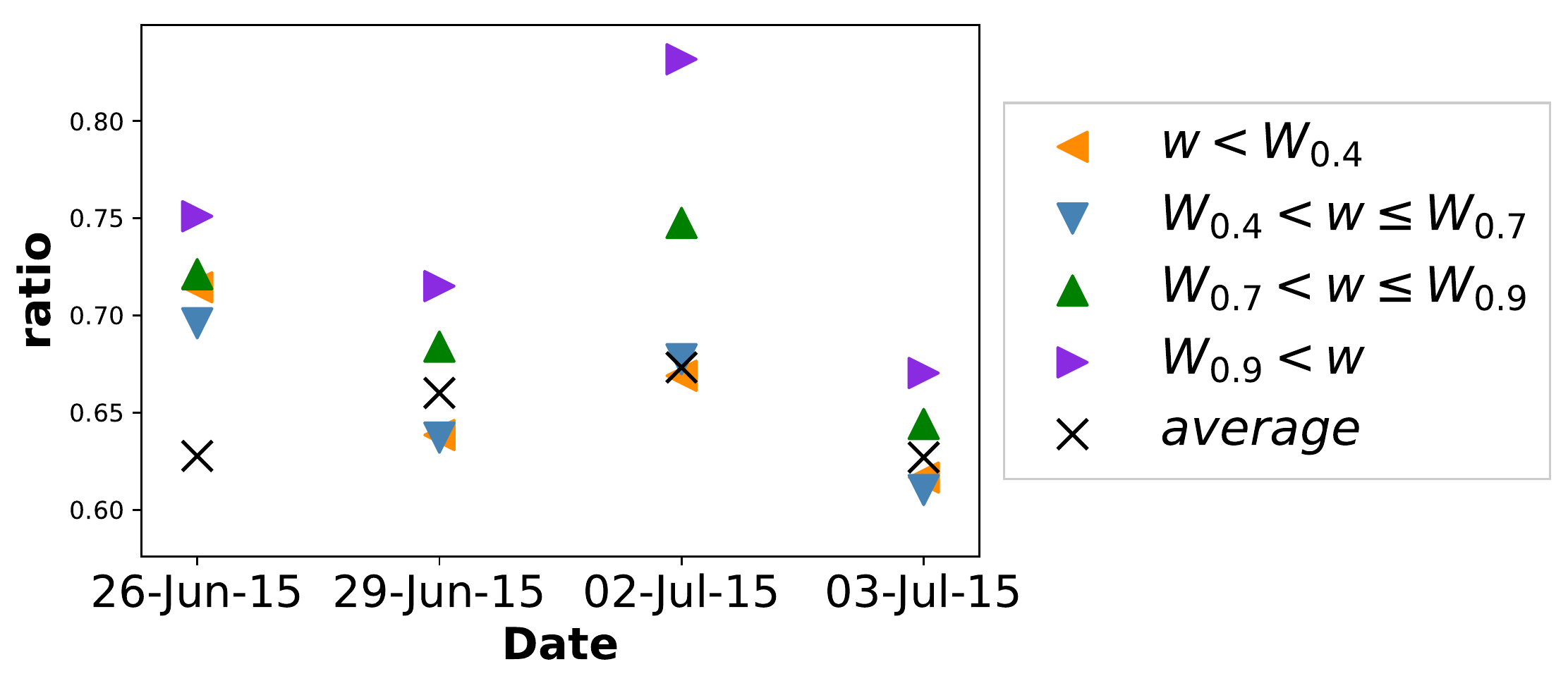}
\caption{\bf{Granger Causality tests for directed links.} \normalfont{According to the descending order of weights, edges in the stock network are divided into 4 groups. $w$ is the edge weight. The 40th quantile, 70th quantile, 90th quantile of weight sequence are denoted as $W_{0.4}$, $W_{0.7}$, $W_{0.9}$, respectively. The y-axis indicator is the proportion of edges that its source is significant a Granger-cause of its target (the pairs of stocks that fail to provide test results due to the data quality have been excluded from denominators of the ratios). The average level is refer to the proportion of any pair of stocks that has Granger Causality relationship.}}\label{fig:granger_edges}
\end{figure}

As the data of mutual fund holdings are snapshots on June 30 2015 and we assume (see discussions in the section of data and Table 2) the holdings would not change too much within a few days, we narrow down the sampling period to four days that around June 30 and all experienced market crash as large amount of stocks reached their down limits. In Fig.~\ref{fig:granger_edges}, the x-axis is the date and the y-axis is the ratio of pairs of stocks that successfully reject the null hypothesis. According to the descending order of weights, edges in our stock network are divided into four groups. $w$ is the edge weight. The 40th quantile, 70th quantile, 90th quantile of weight sequences are denoted as $W_{0.4}$, $W_{0.7}$, $W_{0.9}$. Accordingly, the y-axis indicator is the proportion of edges that its source significantly Granger-cause its target. Moreover, we calculate the proportions of any pair of stocks that has a Granger Causality relationship with respect to their intraday-one-minute percentage changes for each day. The proportions are shown by cross signs. As shown in Fig.~\ref{fig:granger_edges}, during extreme market crisis, for each pair of stocks linked in the present network, the more influential (higher edge weights) the source node, the higher probability of it being a Granger Cause for its target node. 

In addition, Table \ref{tab:5granger} shows that percentage changes of five stocks with highest out-degrees have higher-level of ability to predict the future percentage changes of theirs successors than the average level. This is the case over the four days of market crash. The results imply that, on one hand, both the edge weight and out-degree defined in our network successfully capture the actual economic meaning about the ability of stock price leading. On the other hand, the mutual funds mimicking holding strategies have made certain nodes so `big' in connectivity that they are extremely influential to the others, that is, the too-connected stocks' prices falling provide statistically significant information about their successors' future prices falling. It is worth noting that, though the too-connected stocks are highly connected to each other, usually considered as `rich-club' effect, they do not have such prediction power for each other (see Appendix Fig.~\ref{fig:rich}).

\begin{table}[htbp!]
\centering
\caption{\bf{Granger Causality tests results for edges from nodes with highest out-degrees.}} \label{tab:5granger}

\begin{tabular}{rlrrrr}
  \hline
 Node & 26.Jun & 29.Jun & 2.Jul & 3.Jul \\ 
  \hline
average level & 0.63 & 0.66 & 0.67 & 0.63 \\ 
601318.SH & 0.72 & 0.73 & 0.95 & 0.67 \\ 
601166.SH & 0.79 & 0.77 & 0.94 & 0.62 \\ 
600036.SH & 0.81 & 0.65 & 0.88 & 0.70 \\ 
600016.SH & 0.90 & 0.78 & 0.96 & 0.67 \\ 
600030.SH & 0.81 & 0.84 & 0.96 & 0.71 \\ 
   \hline
\end{tabular}

\caption*{\scriptsize{Note: The average level refers to the proportion of any pair of stocks in the network that has a Granger Causality relationship. The other decimals are the proportions of successors with five most connected stocks being their Granger-causes.}}
\end{table}

Note that, the top five stocks with highest out-degrees as a whole take only 3.6 percent of capitalization of A-shares market in June 2015, which is not a big part, while out-strength of the five nodes in total owns 11.8 percent of the sum of weights of all edges. Thus, it is their connectivity that makes them influential. Recall that the relationship shown in Fig.~\ref{fig:top5_successor}, one conclusion can be drawn that a small downward fluctuation of too-connected stocks will cause their successors drop more in stock market crisis. This reminds us that the overall similar investment ownership strategy results in not a matter of `too-big-to-fail' but `too-connected-to-fail'. While these too-connected stocks behave relatively strong during the market turbulence, the `small' stocks connected to them suffer from sharp fall.

Figs.~\ref{fig:top5_successor}, \ref{fig:granger_edges}, and Table \ref{tab:5granger} also provide evidence on the cascade of price falling from the too-connected stocks to their successors. That is, once the prices of too-connected stocks fall even slightly, the initial losses frighten the huge amount of uninformed investors, thus trigger market panic and many of these investors find themselves needing to sell whatever stocks they hold to runaway from the market. As a result, many other stocks' prices move down and start another round of losses. Accordingly, the too-connected stocks boost the overall risk and lead to a sharp market fall through the spiral. 

The spiral is different from that proposed by Brunnermeier et al.\citep{brunnermeier2009}. Their spiral is primarily because of liquidity shortage that driven by funding problems of speculators, while the present spiral is largely because of the network disassortativeness arising from herding and forced by the premature market full of extremely inexperienced investors who easily affected not only by institutional trading activity but also initial tiny shocks. Though the cascade is quite short (one hop), but it would spread the market panic and amplify the drops of the successors, leading to an abrupt crash of the market.

\section{Discussion}

Collective human behavior has led to several crisis in financial market \citep{preis2013}. Based on investors' strategies and interactions with investment targets, this paper explores the complex issue of too-connect-to-fail in deliberately simplified network design of China stock market in times of market crash. The defined formation mechanism of the stock network sheds light on the importance of market participants' herding behavior in investment. It is shown that mutual fund institutions seek for high diversity across individual investment strategy, however, low diversity across the system as a whole (see Fig.~\ref{fig:heat}) leads to investment concentration in some stocks. Those stocks, which are mainly stocks of large-cap and financial firms (see Appendix Table~\ref{tab:industry} and \ref{tab:style}), become highly connected and the network is therefore disassortative, which in turn has increased the systemic risk. Those who diversified their portfolio a lot are likely to suffer great losses in times of market crash (see Appendix Fig.~\ref{fig:entropy}) and the diversification strategy no longer works because of the high-level of systemic risk. Hence, it is important to reexamine the trade-off between individual investment decision and global efficiency \citep{schweitzer2009}. The severe `herding' among China stock market may be induced by complicated reasons. In particular, with unwillingness to miss out on the bull market ride before the crash happens, investors simply mimic the investment decisions of others \citep{scharfstein1990}. While few studies have integrated herding behavior into the overall stock market risk, our results suggest that the potential of combining extensive investor behavioral factors, especially those irrational ones, could offer a better understanding of the stock market systemic risk, like the too-connected-to-fail risk.

In the knowledge of herding creating too-connected stocks, this paper further stresses sufficient attention should be paid to the possibly more severe issue of too-connected-to-fail. The highly connected stocks are not comparatively `big' enough in stock market value, but has profound influence on other stocks through their connectedness. Our discussion is different from the debate on too-big-to-fail in banking system \citep{sorkin2010, battiston2012}, while in line with consideration of systemic risk \citep{chan-lau2010,leon2011}. Through Granger Causality tests, it is found that even if the too-connected stock is not of complete failure, its bit of falling down is likely to cause market panic to propagate. The panic in turn leads to other stocks' turbulence due to the complex coupled-structure in stock network, where systemic risk grows, and finally exacerbates the market crash. Granger Causality is not necessarily true causality. Nevertheless, from a statistical perspective, the results indicate the predictive power of falling from too-connected stocks, implying that they would be better signals in warning the market at early stages. Given the cascade of stock failures, our findings may neither provide enough support on crash prediction in stock market nor extending to the markets of other countries, but could offer possible mechanism, such as the network structure and the risk contagion channels from stocks with large out-degrees to stocks with small out-degrees, to simulate or rehearse the market crash, which can be triggered by a disturbance of any too-connected stocks.

The herding behavior studied in this paper, admittedly, is not the exclusive pathway for boosting the market crash, but an extremely important one, as it exerts the too-connected-to-fail risk. The presented analysis raises far-reaching implications for the design of public policy. The first and foremost might be that investment restrictions to avoid the emergence of `too-connected-to-fail' should be added, that is to reshape the stock network, making the stock market more robust to shocks \citep{schweitzer2009, roukny2013, bardoscia2017}. Real ecosystems studies suggest that disassortative systems can be more resistant to certain kinds of perturbation \citep{may2008}, while epidemiological works indicate that disassortative networks are more convenient for infection to spread than the assortative ones \citep{gupta1989}. From the evidence of China stock market, regulators should be wary of its disassortative characteristics in networking stocks, with the emergence of too-connected stocks that potentially leading to system-wide spread of sharp fall. In contrast, heterogeneities of investment strategy would turn out to shake off the problem and become a source of stability. Thus, regulations of reshaping network topology, such as systemic risk leverage constraints or transaction tax~\citep{poledna2016}, incentive measures of `minority rewards'~\citep{mann2017}, would be better solutions to lessen the systemic risk. 

Meanwhile, our results have suggested that the too-connected stocks lead and amplify their successors' price failing. This implies that the too-connected stocks as well as stocks on the high-weighted linkages should be the focus of regulation for their capability of amplifying risk. For example, drops of those too-connected stocks at early stages can indeed be warning signals to the market. Intuitively, the too-connected stocks can be ideal targets in saving the market, because their price-rising may inject faith into the market. However, our further exploration demonstrates that unexpectedly, in the growing market, too-connected stocks share the similar rise with their successors and have little predicting power of price increment for their successors (see Appendix Figs.~\ref{fig:1225_time},\ref{fig:scatter2014}, Table~\ref{tab:granger2014}). This helps to explain why the bailout measure, that the People's Bank of China provided cash to brokers to buy stocks of large-cap, state-owned stocks, mostly are nodes with high out-degrees in the network we built, was not efficient enough to rescue the market from drowning in the 2015 market crash.

However, one alternative bailout choice could be reducing the fragility of stocks that are systemically important, likewise limiting the trading activities of stocks that are highly connected to stop them from falling further, to lessen the risk of investor panic cascading to their successors. The `Circuit Breakers Mechanism', a kind of trading curb introduced by China securities surveillance department in the early 2016, however, was a whole market temporary pause. It not only limited the whole market liquidity but also intensified market panic. In contrast, the intentional modularity within the market \citep{haldane2011} may well prevent the cascading failure from spreading around the system. Overall, stock market monitoring policy could obtain some insights from the present approach that illustrates the possible origin for stock market instabilities and systemic risk. 

\section{Conclusion}

In this paper, we propose a new method of networking stocks based on common mutual funds to investigate the market crash in 2015 in China. We find that the China mutual fund institutions herding similar investment strategies on the whole makes some stocks become too-connected. A small downward fluctuation of those stocks contributes to sharp decline of other stocks through the interconnection between investor and equity during market crash, inducing the issue of too-connected-to-fail. From the prospective of systemic risk, policies to deal with the intensity of `too-connected' can be drawn from such network model. Our work could be of interest to both market practitioners and policy makers.

\section{Acknowledgments}
This research was financially supported by National Natural Science Foundation of China (Grant Nos. 71420107025 and 71501005).


\begin{appendices}  

\section{Tables and figures}
\setcounter{figure}{0}   
\renewcommand\thefigure{A\arabic{figure}}   
\setcounter{table}{0}
\renewcommand{\thetable}{A\arabic{table}}

\begin{table}[h]
\caption{\bf{The fraction of features in each category grouped by out-degree to the features of all the stocks.}}\label{tab:5quantile}
\medskip
\begin{tabular}{lcccc}
\hline
Category & total out-degree &total out-strength & total stock market value & sample ratio \\
\hline
 $d=0$ & 0.00 & 0.00 & 0.49 & 0.86  \\
 $0<d\leq D_{0.3}$ & 0.09 & 0.05 & 0.06 & 0.04  \\
 $D_{0.3}<d\leq D_{0.6}$ & 0.26 & 0.16 & 0.12 & 0.04  \\
 $D_{0.6}<d\leq D_{0.9}$ & 0.44 & 0.39 & 0.15 & 0.04  \\
 $D_{0.9}<d$ & 0.22 & 0.40 & 0.18 & 0.01  \\
 \hline
 total & 1.00 & 1.00 & 1.00 & 1.00  \\
\hline
\end{tabular}

\medskip
\normalfont{Note: After excluding nodes with zero out-degree, the 30 quantile, 60 quantile, 90 quantile of out-degree sequence are denoted as $D_{0.3}$, $D_{0.6}$, $D_{0.9}$, respectively. $d$ stands for out-degree. The last column shows a benchmark proportions of each stock category. Column 1 and 2 show that the most connected stocks are influential in both out-degree and out-strength. The sum of out-strength of nodes with top out-degrees represents 40 $\%$ of the total edge weights in network. In terms of market value, which is a real world characteristic of stock, the group of `$D_{0.9}<d$' owns only 0.18 proportion of market value to total 2709 stocks.}
\end{table}

\begin{table}[h!]
\caption{\bf{Numbers (proportions) of stocks by sector in different out-degree categories.}}\label{tab:industry}

\resizebox{\textwidth}{!}{\begin{tabular}{rllllll}
  \hline
category & $d=0$ & $0<d\leq D_{0.3}$ & $D_{0.3}<d\leq D_{0.6}$ & $D_{0.6}<d\leq D_{0.9}$ & $D_{0.9}<d$  & total  \\ 
  \hline
finance & 6 & 1 & 10 & 16 & 16 & 49 \\ 
 & (0.122) & (0.02) & (0.204) & (0.327) & (0.327) & (1) \\ 
information technology  & 175 & 24 & 18 & 23 & 7 & 247 \\ 
 & (0.709) & (0.097) & (0.073) & (0.093) & (0.028) & (1) \\ 
transportation & 71 & 1 & 2 & 2 & 3 & 79 \\ 
 & (0.899) & (0.013) & (0.025) & (0.025) & (0.038) & (1) \\ 
manufacturing & 1432 & 70 & 61 & 44 & 8 & 1615 \\ 
 & (0.887) & (0.043) & (0.038) & (0.027) & (0.005) & (1) \\ 
communication and culture & 36 & 2 & 3 & 8 & 1 & 50 \\ 
 & (0.72) & (0.04) & (0.06) & (0.16) & (0.02) & (1) \\ 
agriculture & 39 & 2 & 3 & 1 & 0 & 45 \\ 
  & (0.867) & (0.044) & (0.067) & (0.022) & (0) & (1) \\ 
 construction & 51 & 4 & 2 & 3 & 0 & 60 \\ 
   & (0.85) & (0.067) & (0.033) & (0.05) & (0) & (1) \\ 
  real estate & 117 & 2 & 0 & 2 & 2 & 123 \\ 
  & (0.951) & (0.016) & (0) & (0.016) & (0.016) & (1) \\ 
 retailing & 119 & 6 & 4 & 6 & 0 & 135 \\ 
  & (0.881) & (0.044) & (0.03) & (0.044) & (0) & (1) \\ 
 electricity, gas, water & 77 & 2 & 4 & 1 & 0 & 84 \\ 
  & (0.917) & (0.024) & (0.048) & (0.012) & (0) & (1) \\ 
 service industry & 88 & 1 & 6 & 7 & 2 & 104 \\ 
  & (0.846) & (0.01) & (0.058) & (0.067) & (0.019) & (1) \\ 
 extractive industry & 65 & 1 & 3 & 1 & 0 & 70 \\ 
  & (0.929) & (0.014) & (0.043) & (0.014) & (0) & (1) \\ 
 other & 42 & 1 & 1 & 2 & 0 & 46 \\ 
  & (0.913) & (0.022) & (0.022) & (0.043) & (0) & (1) \\ 
  \hline
Sample size & 2318 & 117 & 117 & 116 & 39 & 2707 \\ 
& (0.856)& (0.043)& (0.043)& (0.042)& (0.014)& (0.999)\\
   \hline
\end{tabular}}

\medskip
\normalfont{Note: The values in brackets are proportions of stocks in different categories. The ratios in the last row called `sample size' can be viewed as benchmark. There are two stocks without industry label and are excluded in this table. The group of nodes with high out-degrees is dominated by stocks from financial sector (see column 5 in finance row). }
\end{table}

\begin{table}[th!]
\caption{\bf{Proportions of stocks by equity style in different out-degree categories.}}\label{tab:style}
\medskip
\resizebox{\textwidth}{!}{\begin{tabular}{rcccccc}
  \hline
category & $d=0$ & $0<d\leq D_{0.3}$ & $D_{0.3}<d\leq D_{0.6}$ & $D_{0.6}<d\leq D_{0.9}$ & $D_{0.9}<d$  & sample ratio \\ 
  \hline
large-cap-value & 0.06 & 0.18 & 0.14 & 0.15 & 0.31 & 0.08 \\ 
large-cap-balance & 0.09 & 0.23 & 0.26 & 0.29 & 0.41 & 0.12 \\ 
large-cap-growth & 0.05 & 0.22 & 0.30 & 0.47 & 0.28 & 0.09 \\ 
mid-cap-value & 0.09 & 0.04 & 0.01 & 0.02 & 0.00 & 0.08 \\ 
mid-cap-balance & 0.15 & 0.14 & 0.08 & 0.03 & 0.00 & 0.14 \\ 
mid-cap-growth & 0.09 & 0.14 & 0.19 & 0.03 & 0.00 & 0.09 \\ 
small-cap-value & 0.11 & 0.00 & 0.00 & 0.00 & 0.00 & 0.09 \\ 
small-cap-balance & 0.24 & 0.02 & 0.01 & 0.00 & 0.00 & 0.21 \\ 
small-cap-growth & 0.12 & 0.02 & 0.01 & 0.00 & 0.00 & 0.10 \\ 
\hline
total & 1 & 1 & 1 & 1 & 1 & 1 \\ 
   \hline
\end{tabular}}

\medskip
\normalfont{Note: The last column is the benchmark proportions by equity style. The most connected nodes ($D_{0.9}<d$ column) are mainly large-cap stocks. Most nodes with zero out-degree are small-cap stocks ($d=0$ column).}
\end{table}

\begin{table}[h]

\caption{\bf{Granger Causality test results for network in December 2014.}}\label{tab:granger2014}
\medskip

\begin{tabular}{lcc}
\hline
\multicolumn{2}{c}{Granger Causality test for top five nodes with highest out-degrees and their successors}\\
  \hline
node & proportion of successors with the node being their Granger-causes  \\ 
\hline
601318.SH & 0.240 \\ 
601166.SH & 0.285 \\ 
000002.SH & 0.262 \\ 
600036.SH & 0.266 \\ 
600000.SH & 0.288 \\ 
 \hline 
\hline
\multicolumn{2}{c}{Granger Causality tests for directed links}\\
  \hline
edges groups & proportion of edges that its source is a Granger-cause of its target  \\ 
\hline
average level & 0.146 \\
$0<w\leq W_{0.4}$& 0.180 \\ 
$W_{0.4}<w\leq W_{0.7}$ & 0.197 \\ 
$W_{0.7}<w\leq W_{0.9}$ & 0.217 \\ 
$w\leq W_{0.9}$ & 0.277 \\ 
   \hline 
\end{tabular}

\medskip
\normalfont{Note: The average level refers to the proportion of any pair of stocks in the Dec-2014 network that has a Granger Causality relationship. The proportions are much lower than that presented in Fig.~\ref{fig:granger_edges} and Table~\ref{tab:5granger} in the main text, implying the weak predicting power of the source nodes for their target nodes in the Dec-2014 network. Nevertheless, the groups of edges with higher weights have higher proportions than lower weights groups, and the most connected stocks as well as the stocks linked together have higher test-passing ratio that the average level, indicating the edges defined in our networking approach successfully capture the actual economic meaning about the ability of stock price leading. The details on how and why the Dec-2014 network is built can be found in main text.}
\end{table}

\begin{table}[htbp!]
\caption{\bf{The five mutual fund institutions that invest the most per stock in the `$D_{0.9}<d$' group.}}\label{tab:five_companies}
\medskip
\resizebox{\textwidth}{!}{\begin{tabular}{rcccc}
  \hline
\shortstack{Mutual fund \\institution} & \shortstack{The average \\holding market \\value per stock in \\`$D_{0.9}<d$' group\\(unit: 1000 CNY)} & \shortstack{The number of \\stocks held in \\`$D_{0.9}<d$' \\group} & \shortstack{Proportion of \\stocks held in\\ `$D_{0.9}<d$' \\group to the \\number of all \\stocks in the group} & \shortstack{The number \\of stocks \\held in total} \\ [3pt]
  \hline
China AMC & 1157086 &  40 & 1.00 & 1208 \\ 
Fullgoal Fund  & 938770 &  39 & 0.97 & 1269 \\ 
Penghua Fund & 864474 &  40 & 1.00 & 1104 \\ 
E Fund & 719693 &  40 & 1.00 & 910 \\ 
SWS MU Fund & 557249 &  36 & 0.90 & 892 \\ 
  \hline
\end{tabular}}
\end{table}

\newpage

\begin{figure}[ht!]
\centering
\includegraphics[width= 0.5\linewidth]{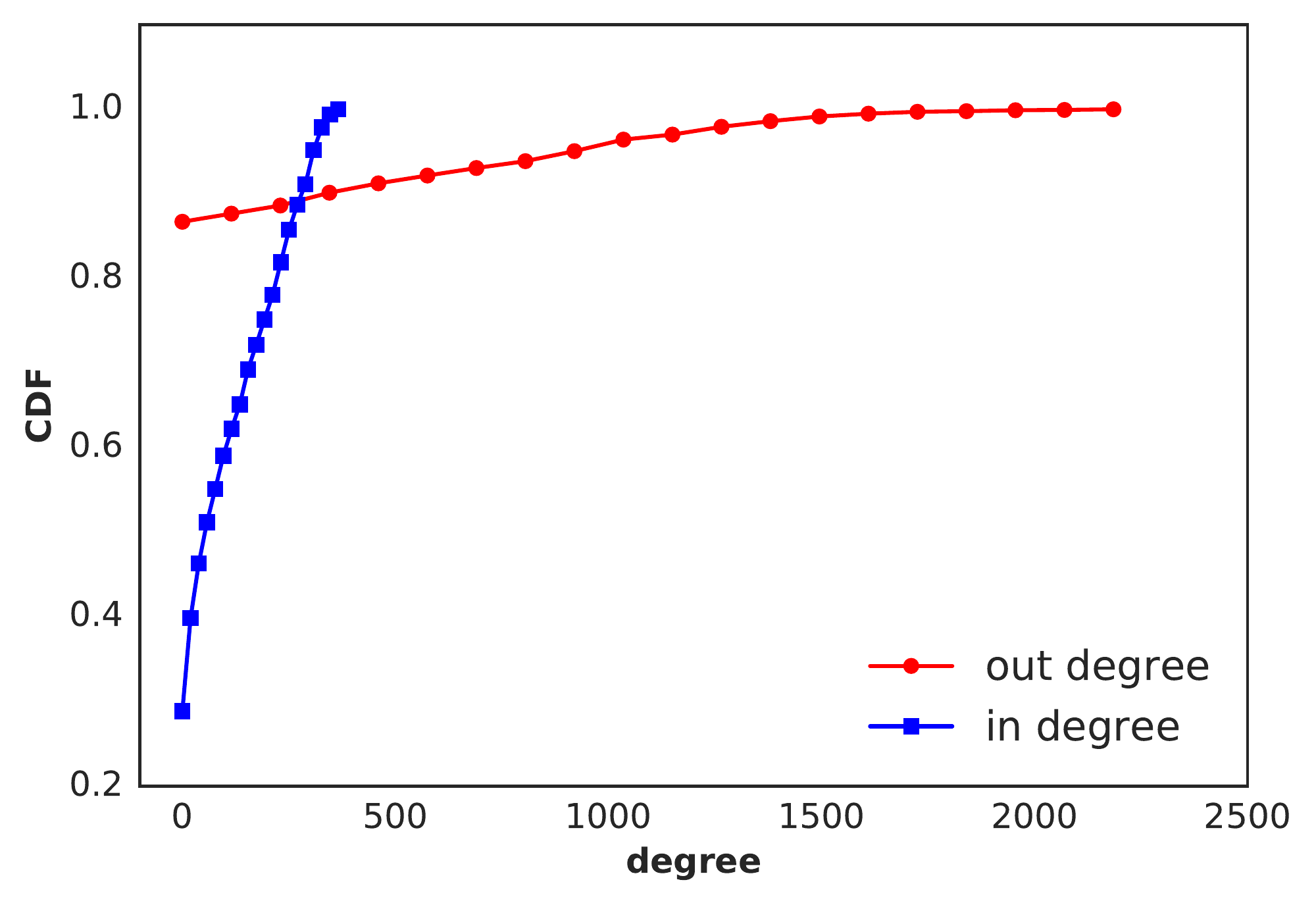}
\caption{\bf{Degree cumulative distribution function.} \normalfont{The out-degree strongly leans towards zero. In fact, there are 2319 out of 2709 nodes with out-degree equal to zero, and nodes with out-degrees higher than 1500 are left sporadically. The in-degree uniformly distributes in the range of 0 to 400, reflecting that each stock are roughly equally affected by other stocks. The average out-degrees is about 115.}}
\label{fig:degree_cdf}
\end{figure}

\begin{figure}[ht!]
	\begin{minipage}{0.5\linewidth}
		\centering
		{\footnotesize (a)}
		\includegraphics[width =0.95\linewidth]{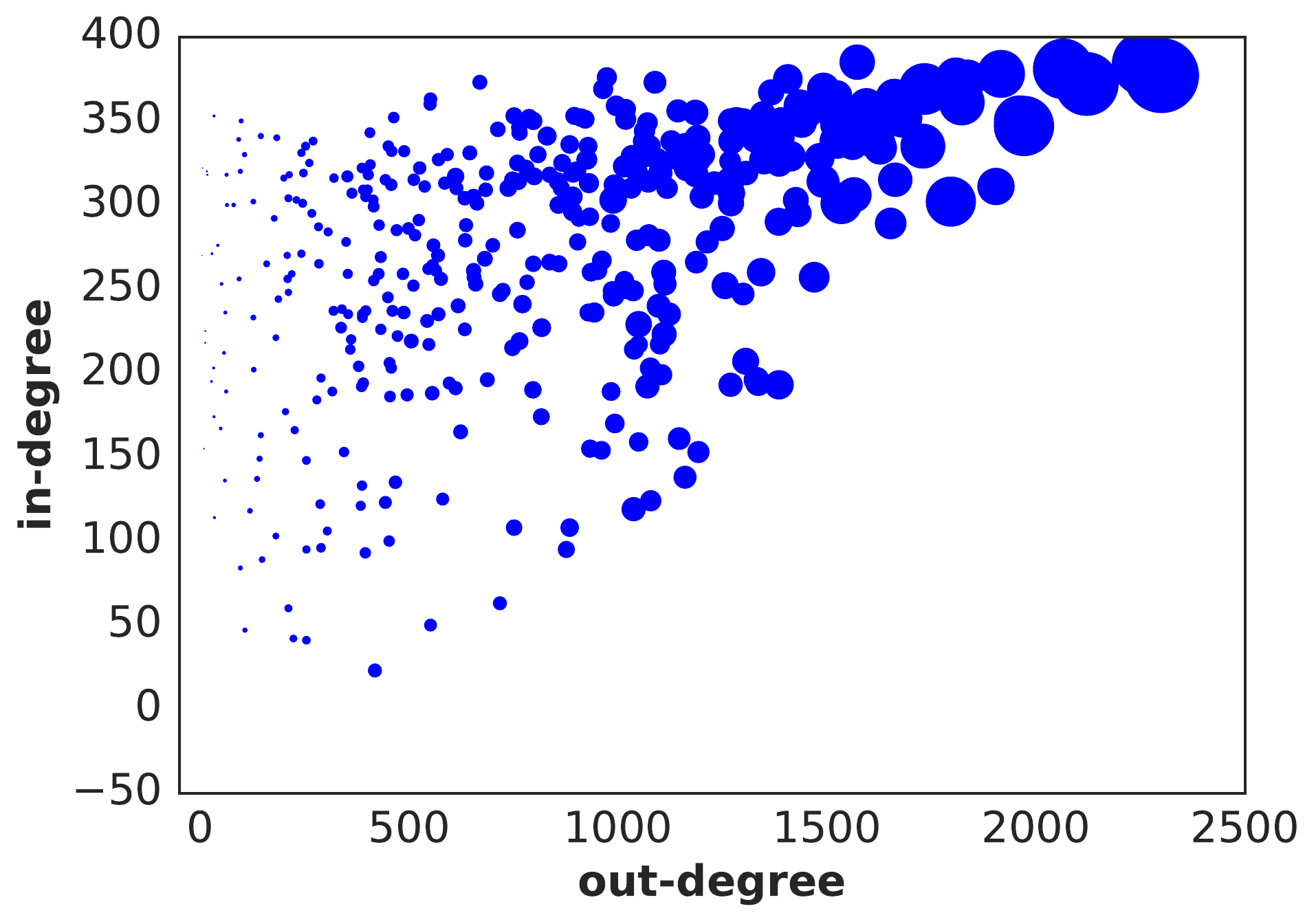}
	\end{minipage}\hfill
	\begin{minipage}{0.5\linewidth}
		\centering
		{\footnotesize (b)}
		\includegraphics[width =0.95\linewidth]{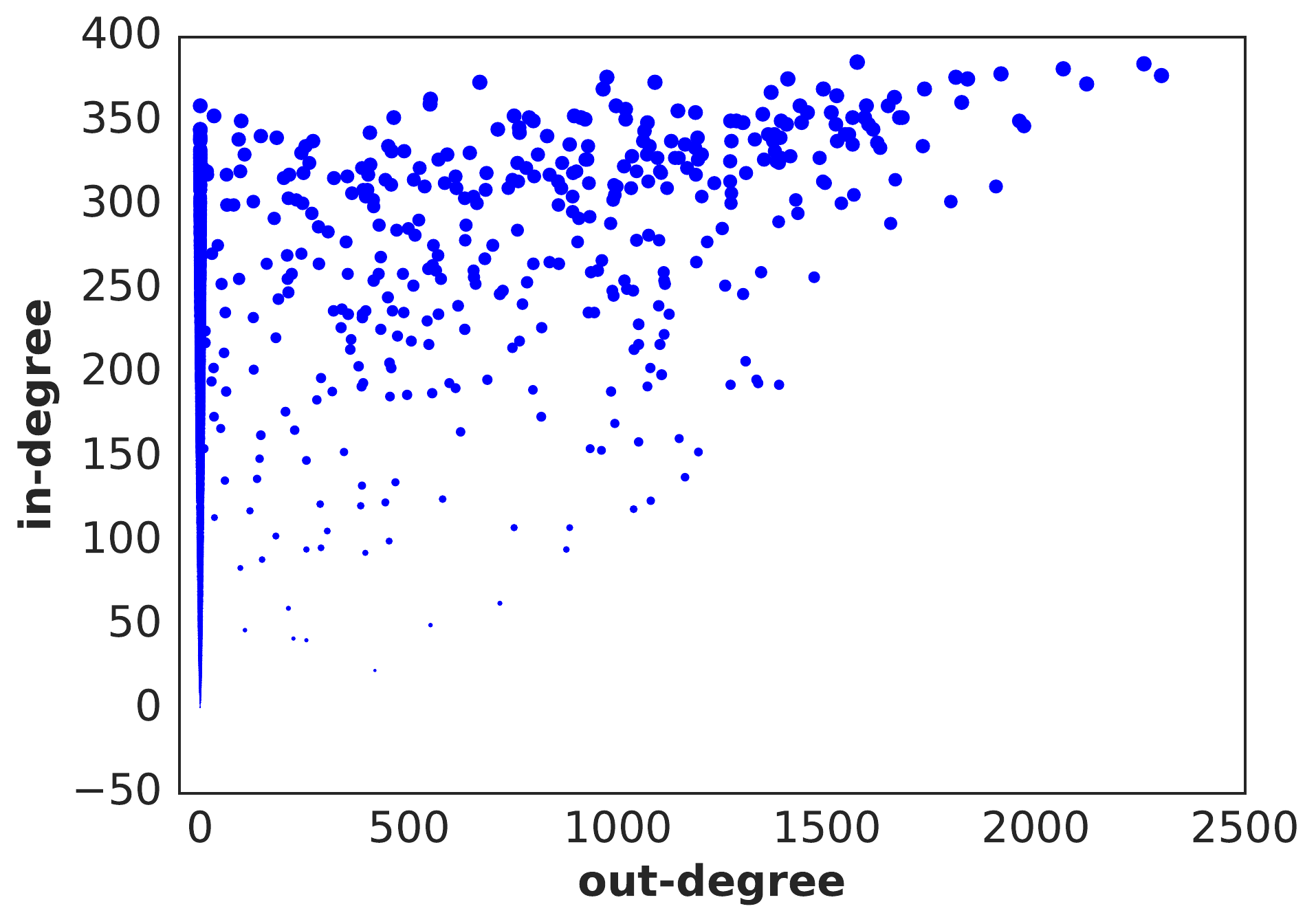}
	\end{minipage}
\caption{\bf{The correlation between the out-degree and in-degree.} \normalfont{The bubble size in (a) and (b) is respectively proportional to out-strength and in-strength of the node. Nodes with out-degree equal to 0 have zero out-strength and they are naturally ignored in the plot (a). (a) shows that a few stocks are large in out-degree as well as out-strength. The out-strength is highly correlated to out-degrees as the bubble size increases with the out-degree. (b) exhibits that in-degree is weakly correlated to in-strength, and nodes with top in-degrees are not all top out-degrees nodes. Because the existence of a number of nodes with out-degree equal to 0, there's a ribbon on the left, showing that when a node has no out-degree, it may have in-degrees. }}\label{fig:bubble}
\end{figure}

\begin{figure}[ht!]
\centering
\includegraphics[width= 3in]{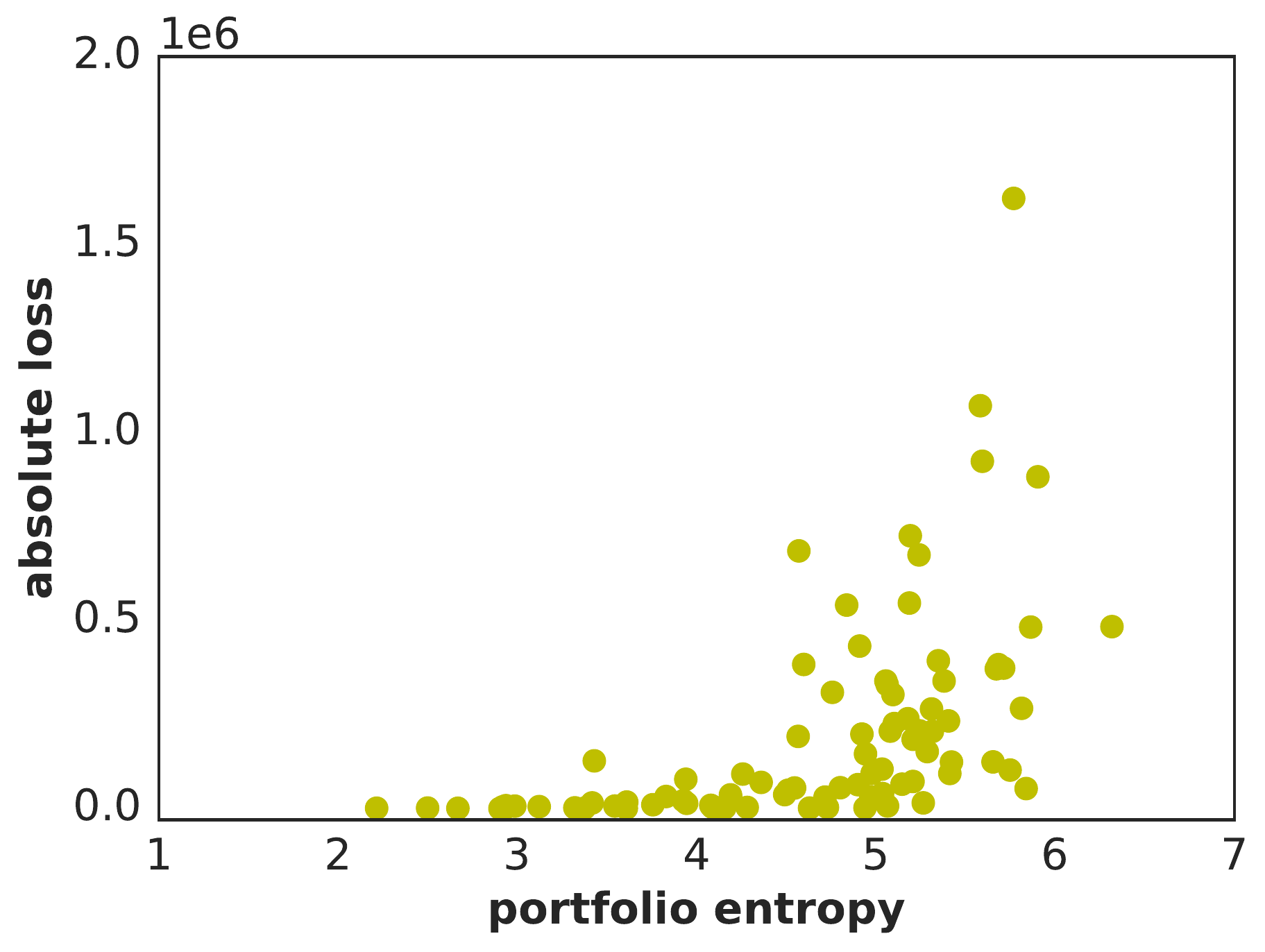}
\caption{\bf{The portfolio entropy of 87 mutual fund institutions and their absolute loss on June 26, 2015.} \normalfont{June 26 is the day with most stocks declining to their down limits. The information entropy of investment portfolio for each mutual fund institution is $\varepsilon_j=-\sum_{i=1}^n\frac{h_{ji}}{s_j}log\frac{h_{ji}}{s_j}$, where $h_{ji}$ is the capitalization of investor $j$ holds for stock $i$, $s_j$ is the total market value of all stocks investor $j$ holds. Accordingly greater $\varepsilon_j$ indicates higher diversity of $j$'s investment strategy. There are $n$ stocks available in the market. $h_{ji}=0$ when investor $j$ has no investment in stock $i$. The absolute loss for investor $j$ is $l_j=|\sum_{i=1}^nh_{ji}d_i|$, where $d_i$ is the net percentage change of stock $i$ on Jun 26. The entropy-based portfolio evaluation suggests that the individual goal of maximizing the portfolio diversification, which results in high heterogeneity of the entire market, failed in times of market crash. }}
\label{fig:entropy}
\end{figure}

\begin{figure}[ht!]
	\begin{minipage}{0.5\linewidth}
		\centering
		{\footnotesize (a)}
		\includegraphics[width =0.95\linewidth]{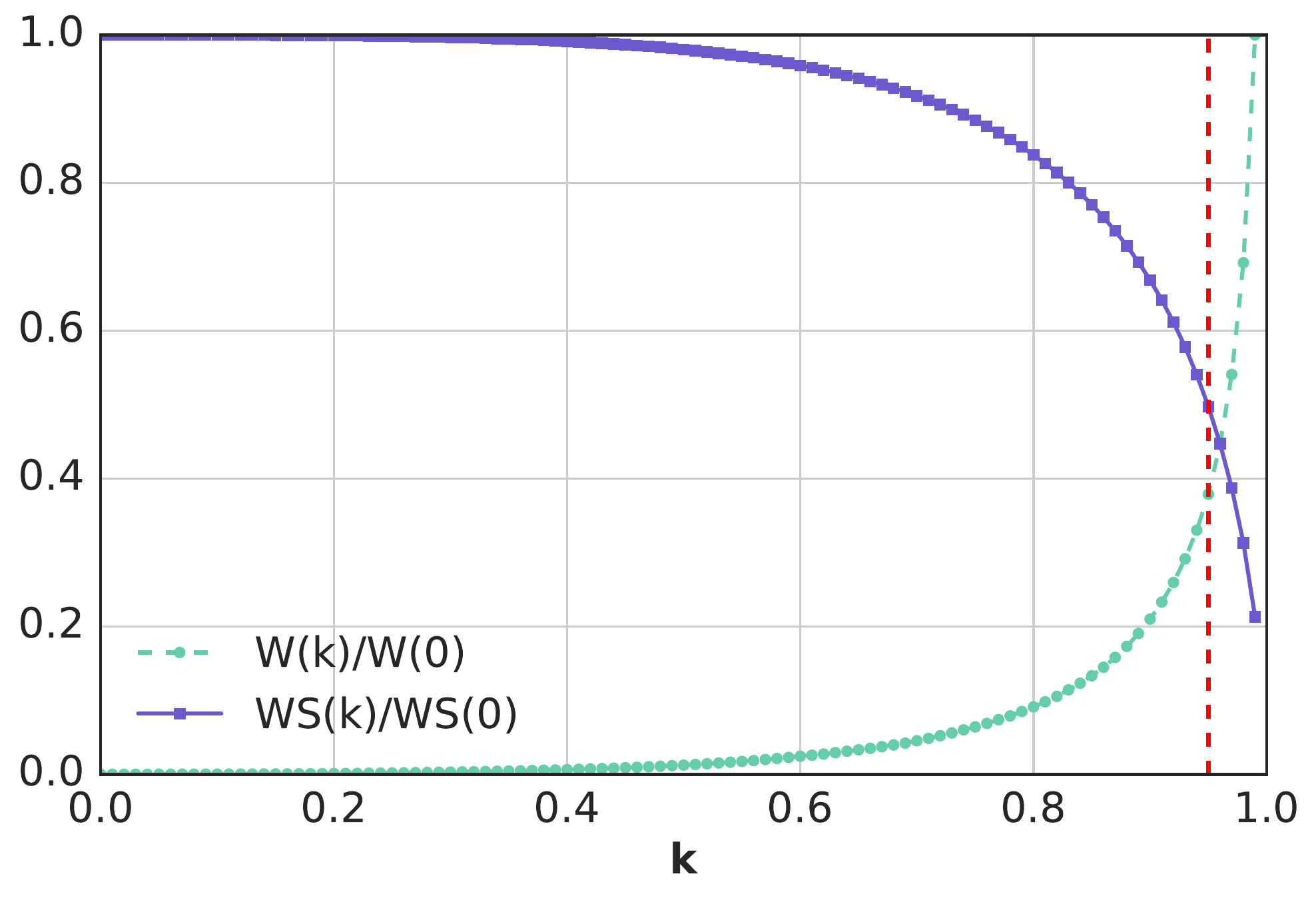}
	\end{minipage}\hfill
	\medskip
	\begin{minipage}{0.5\linewidth}
		\centering
		{\footnotesize (b)}
		\includegraphics[width =0.95\linewidth]{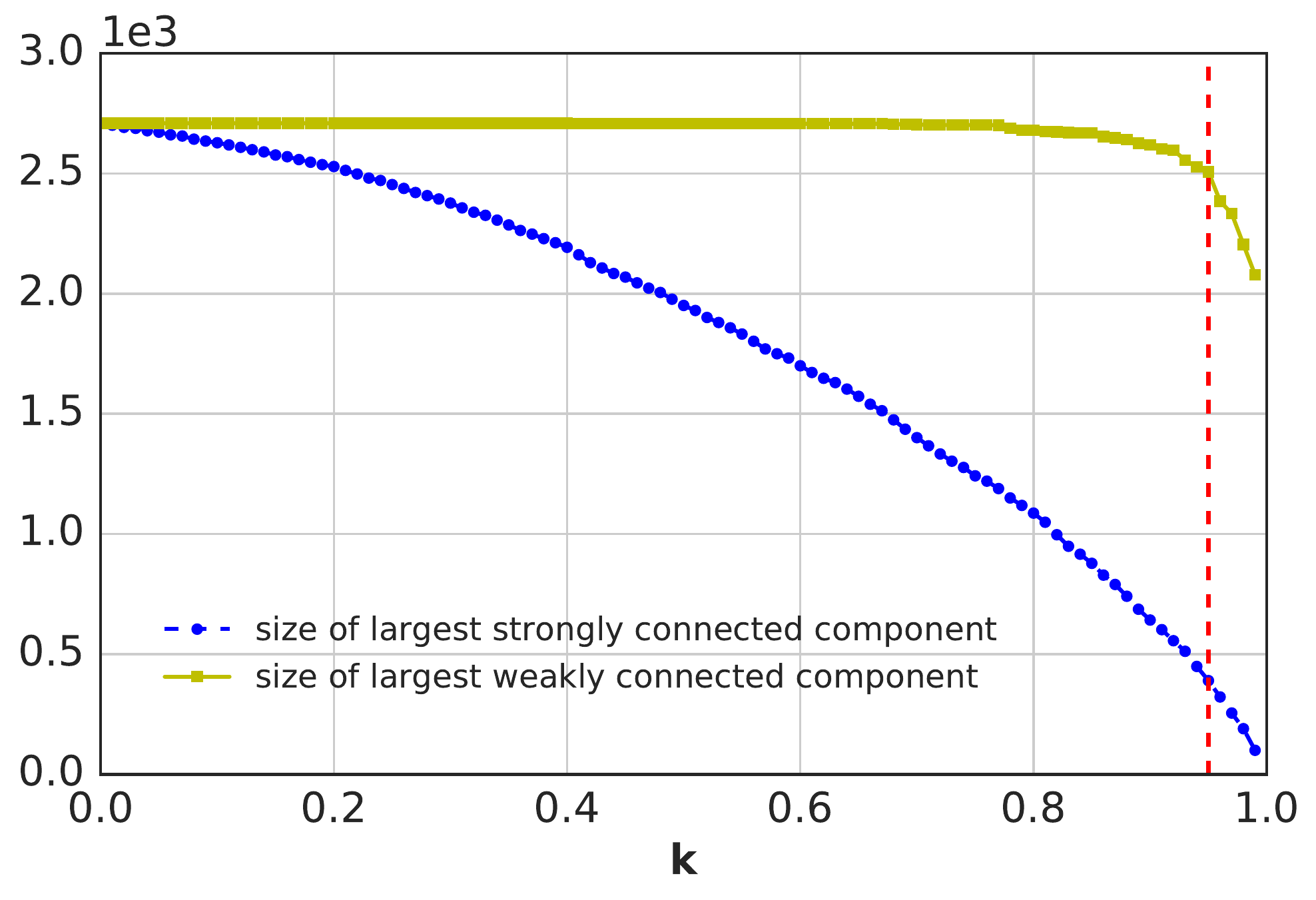}
	\end{minipage}

\caption{\bf{Measures of network as functions of the parameter $k$.} \normalfont{A parameter denoted as $k$ is range from 0 to 1. For each $k$, we delete edges that have weight less than the $100*k$th-quantile of the weight sequence in the non-filtering network, and then calculate structural characteristics of the updated network. In (a), $W(k)$ is the $100*k$th-quantile weight of the non-filtering network. $W(0)$ is the largest edge weight of the non-filtering network. $WS(k)$ is the sum of the weights of all the edges in the updated network. $WS(0)$ is the sum of the weights of all the edges in the non-filtering network. To retain the informative edges (edges with higher weight) as well as to keep the network connectivity (the size of largest weakly connected component) with little change, the value $k=0.95$ is the threshold used in the paper (red line of dashes), that is, delete edges with weights lower than the 95th percentile of edge weights sequence. The established network of the stock market possesses 2709 nodes and 313307 edges after necessary filtering. }}\label{fig:filter}
\end{figure}

\begin{figure}[ht!]
		\centering
		\includegraphics[width =0.5\linewidth]{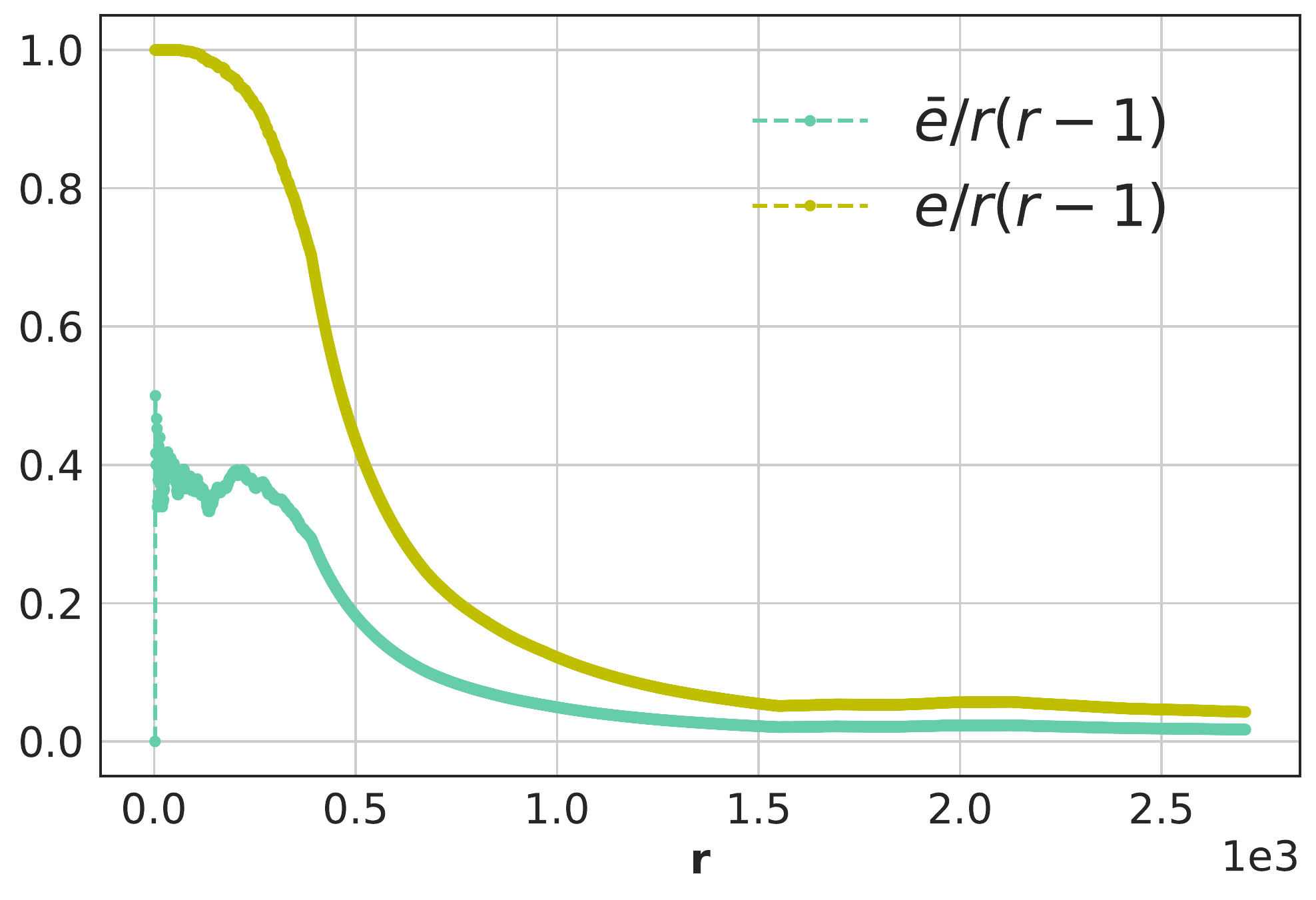}
\caption{\bf{Rich-club effect.} \normalfont{The x-axis denotes the number of top $r$ nodes of high out-degrees. The number of edges in the subgraph constituted by the top $r$ nodes of high out-degrees is denoted as $e$. $r(r-1)$ measures the maximum number of edges that $r$ nodes could have in a directed graph. The number of edges in the subgraph constituted by top $r$ nodes of high out-degrees that passes the granger test is denoted as $\bar{e}$ and accordingly $\bar{e}/r(r-1)$ is the extent of interactive influence among the top $r$ nodes of high out-degrees.}}\label{fig:rich}
\end{figure}

\begin{figure}[ht!]
	\begin{minipage}{0.5\linewidth}
		\centering
		{\footnotesize (a)}
		\includegraphics[height = 3.5in]{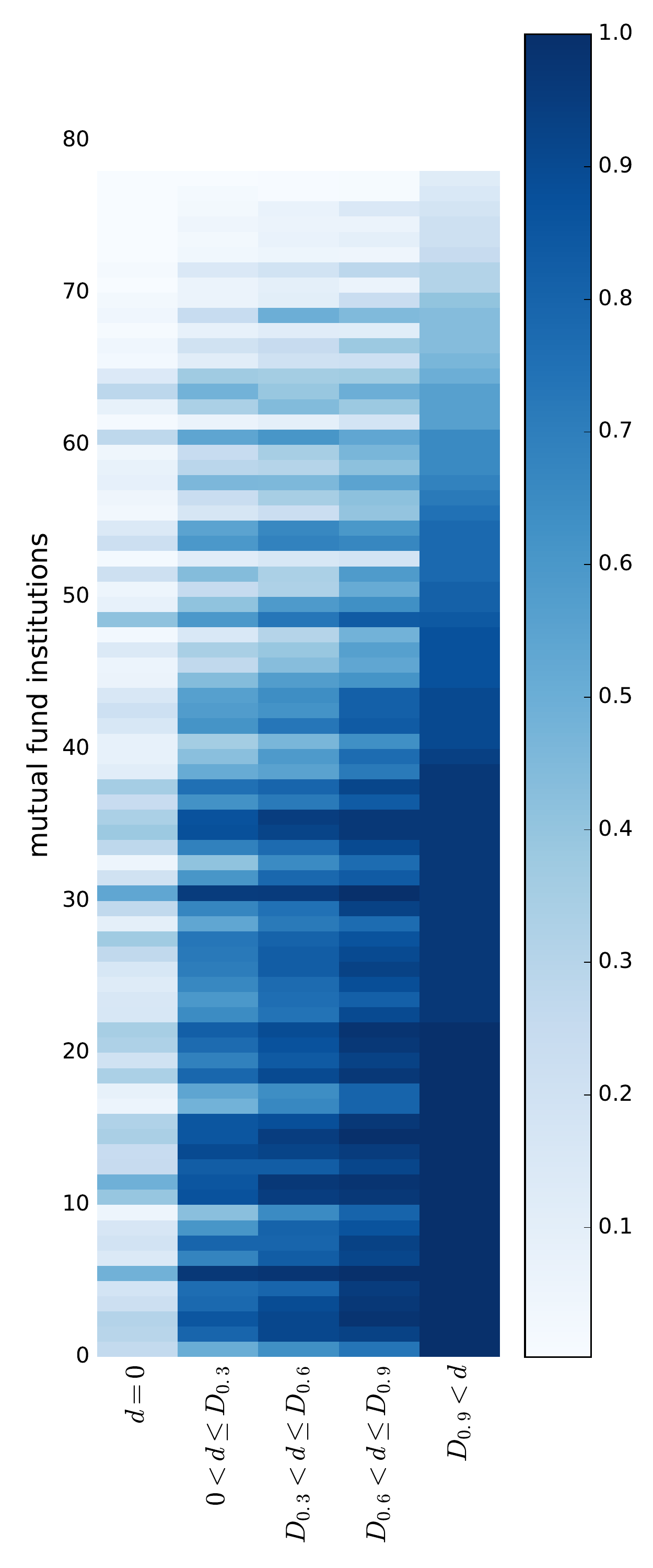}
	\end{minipage}\hfill
	\begin{minipage}{0.5\linewidth}
		\centering
		{\footnotesize (b)}
		\includegraphics[height = 3.5in]{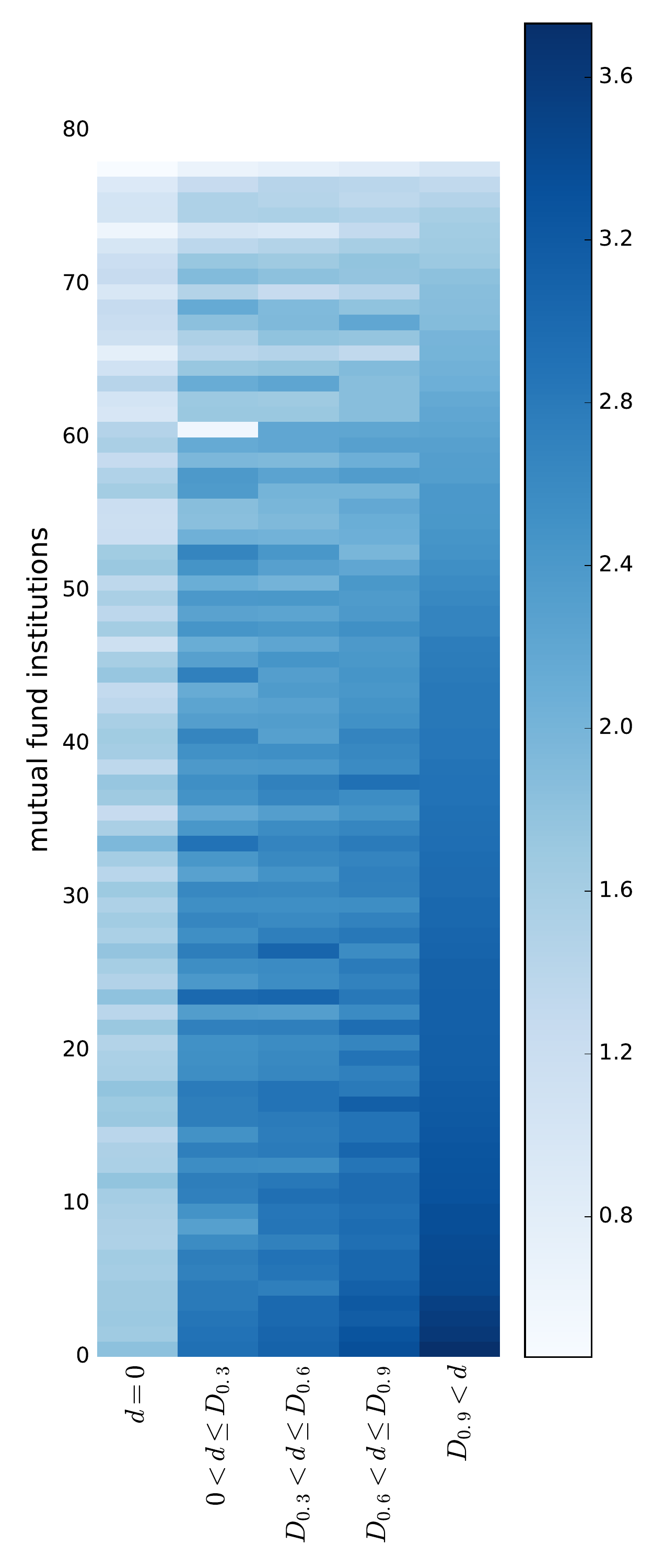}
	\end{minipage}\hfill
\caption{\bf{Mutual fund institutions' investment patterns in December 2014.}
\normalfont{The details on how and why the Dec-2014 network is built can be found in main text. The mutual fund institutions are labeled from 1 to 87 vertically. Every institution has one horizontal color bar showing its preference of investments. (a) Each grid presents the number of stocks one institution hold divided by the number of stocks in the corresponding out-degree category. (b) Each grid presents the average market value per stock one institution hold in the corresponding out-degree category. After excluding nodes with zero out-degree, the 30 quantile, 60 quantile, 90 quantile of out-degree sequence are denoted as $D_{0.3}$, $D_{0.6}$, $D_{0.9}$, respectively. $d$ stands for the out-degree of nodes. The figures show similar gradient color from low out-degree category to high out-degree category, implying that most mutual fund institutions have the preference on stocks with large out-degrees. The relevant one-tailed, paired-samples $t$ tests confirm the significance of this kind of preference. The overall pattern is much similar with Fig.~\ref{fig:heat} in the main text, indicating the herding in investment we revealed is consistent and independent to the market performance.}}\label{fig:dec2014herd}
\end{figure}

\begin{figure}[ht!]
\centering
\includegraphics[width=\linewidth]{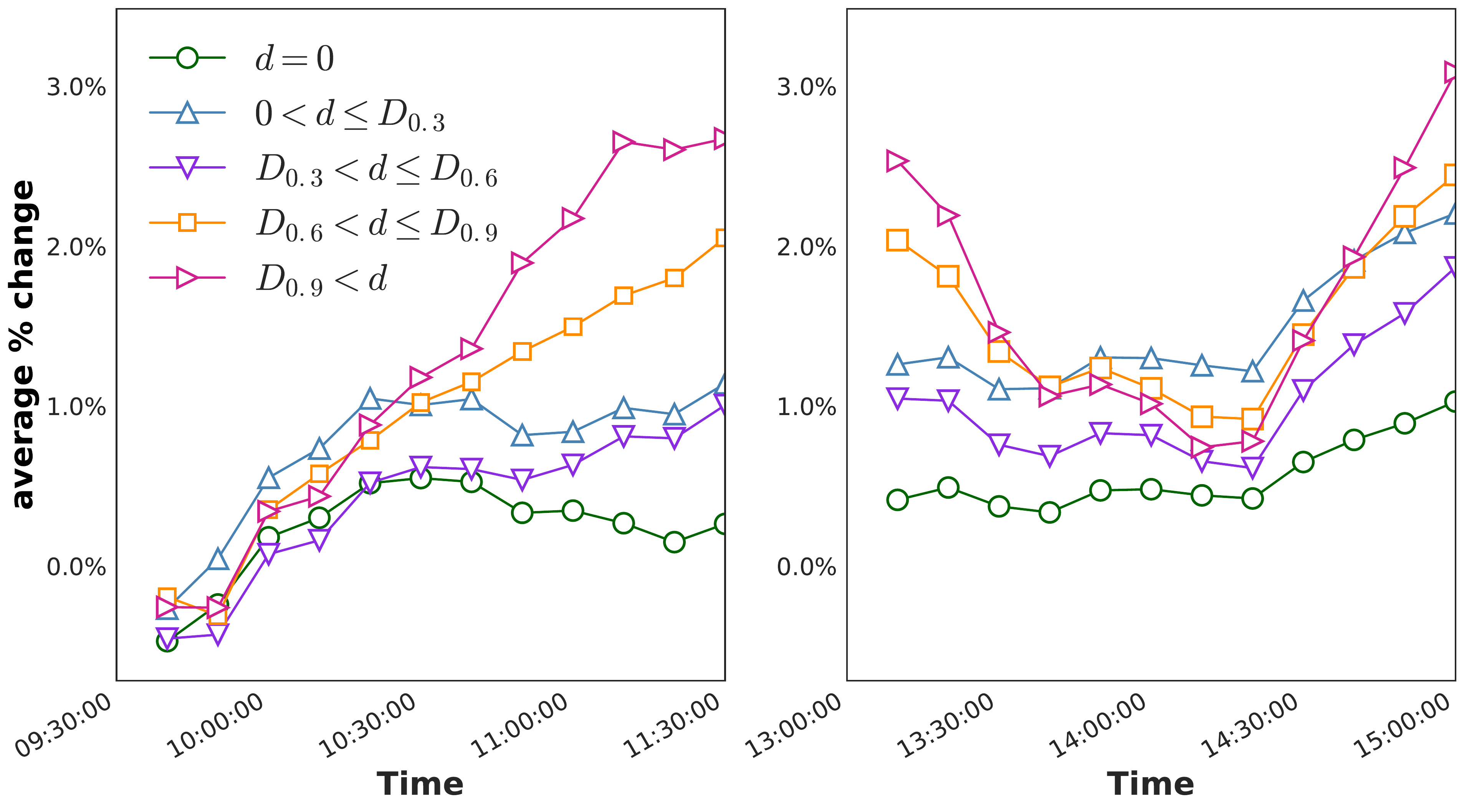}
\caption{\bf{Average percentage changes of five groups' stocks on December 25, 2014.} \normalfont{The details on how and why the Dec-2014 network is built can be found in main text. Intraday percentage changes are computed from data that downloaded from Thomson Reuters' Tick History. The five groups of stocks experienced similar positive percentage changes when the market jump.}}\label{fig:1225_time}
\end{figure}

\begin{figure}[ht!]
\centering
\includegraphics[width=.6\linewidth]{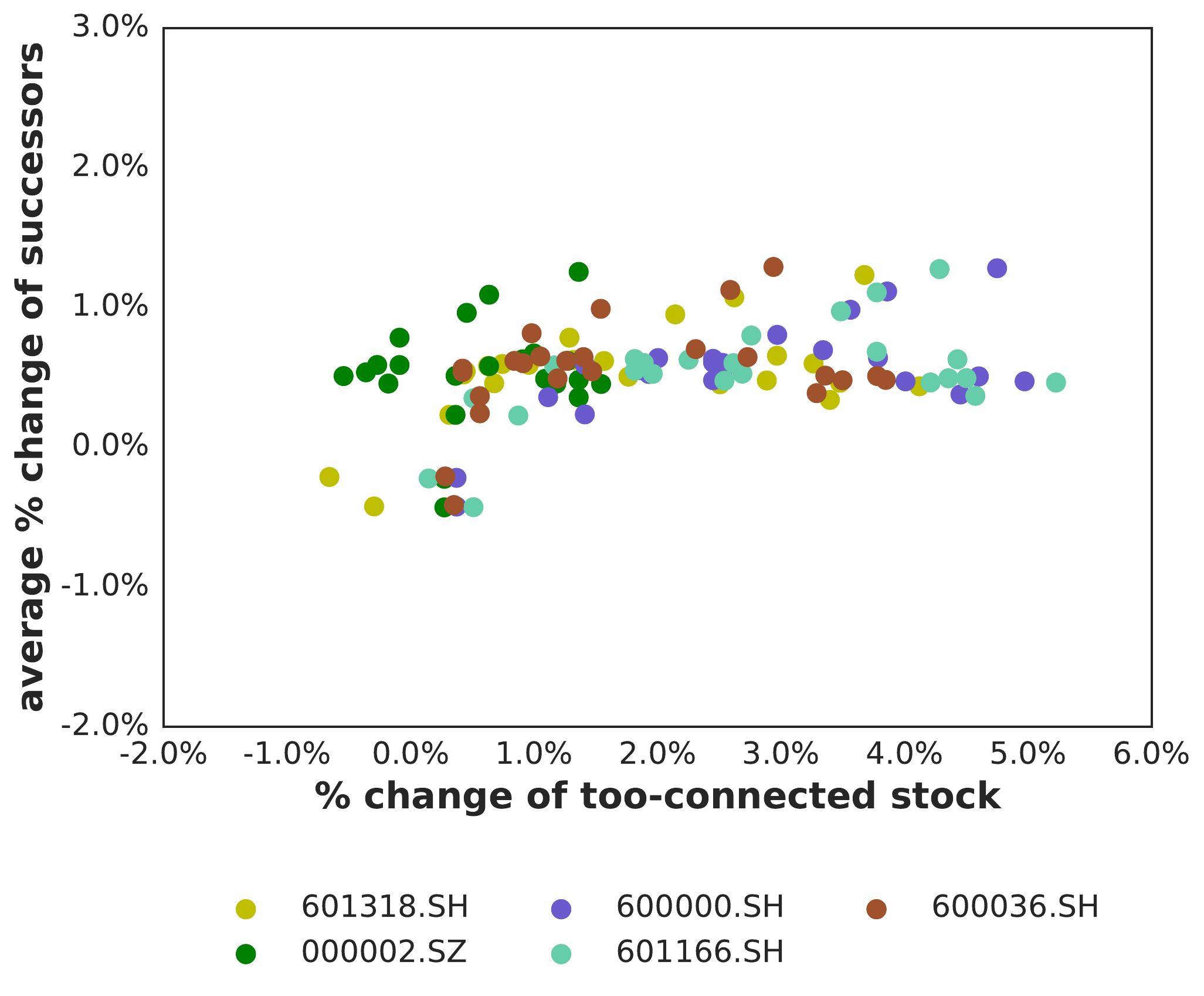}
\caption{\bf{Correlations between `too-connected' stocks' percentage changes and their successors' in December 2014.} \normalfont{The details on how and why the Dec-2014 network is built can be found in main text. The average percentage change of a stock in every ten minutes on December 25, 2014 is used as its market performance, while each highly connected stock has a single percentage changes time series that consists of 24 points and corresponding successors are specified by an averaged percentage change. Round circles refer to stocks with highest out-degrees in our network. They all have more than 1900 successors. Three of them are among the top five nodes with highest out-degrees in network of June 2015.}}\label{fig:scatter2014}
\end{figure}

\begin{figure}[ht!]
	\begin{minipage}{\linewidth}
		\centering
		\includegraphics[width =0.5\linewidth]{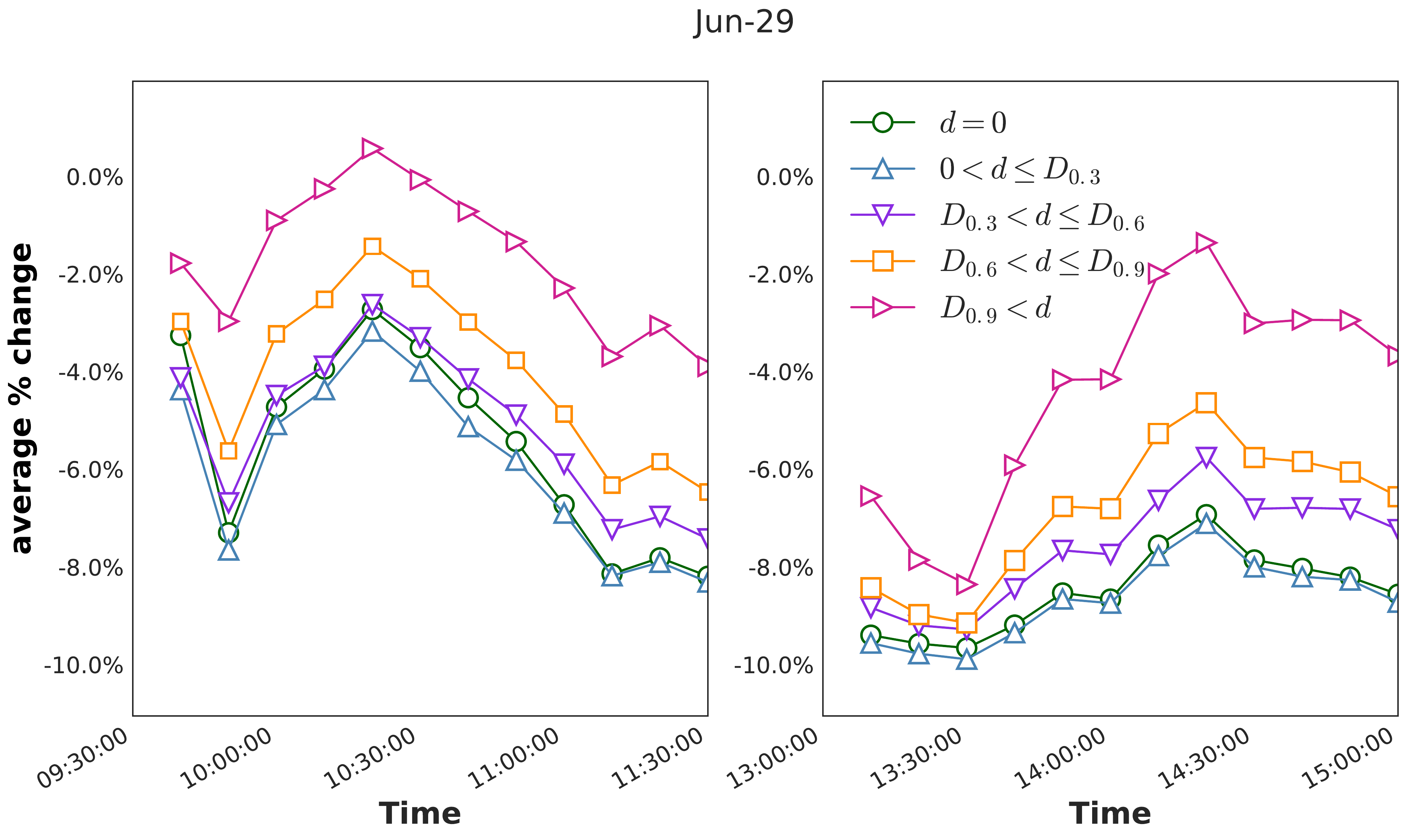}
	\end{minipage}
	\medskip
	\begin{minipage}{\linewidth}
		\centering
		\includegraphics[width =0.5\linewidth]{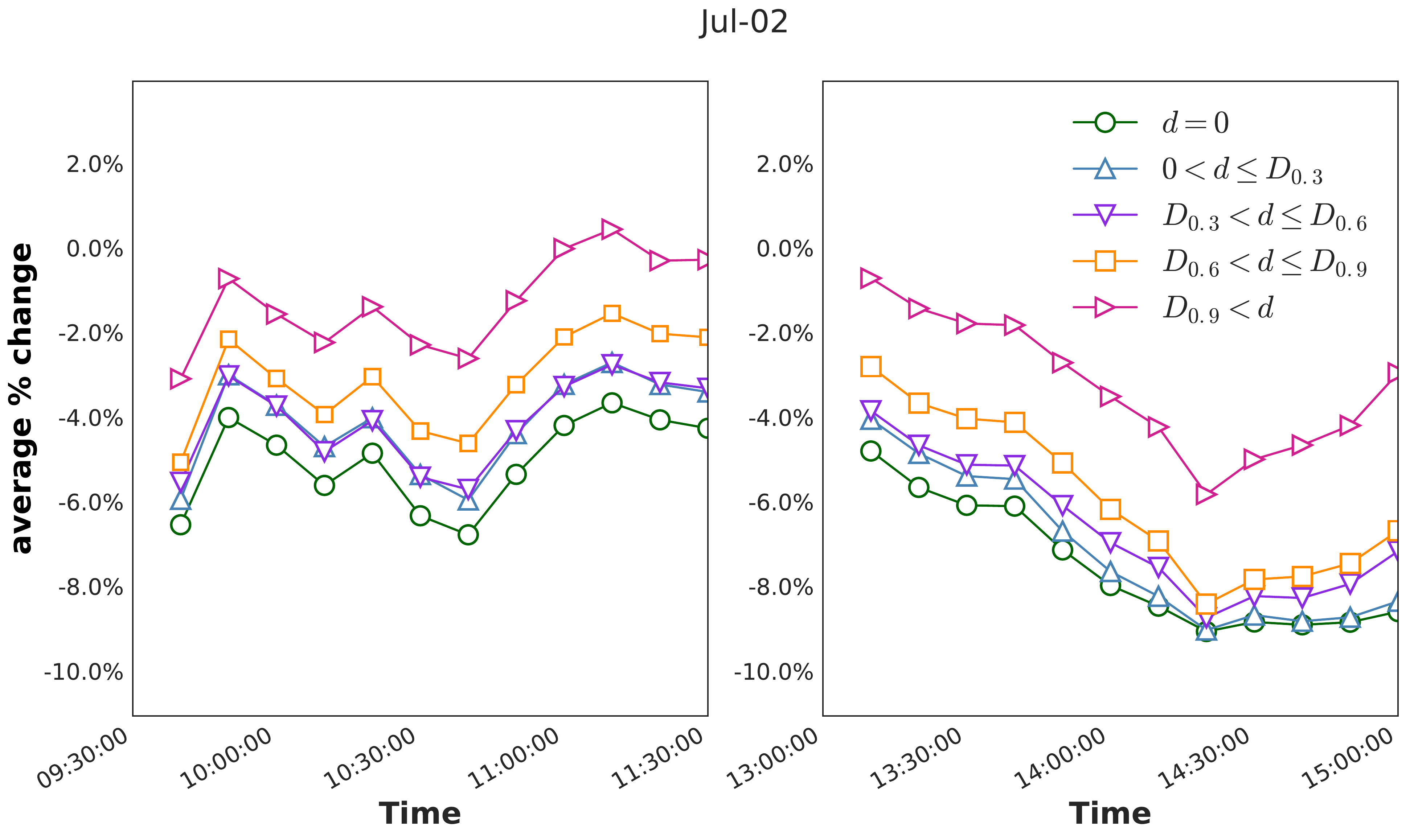}
	\end{minipage}
	\begin{minipage}{\linewidth}
		\centering
		\includegraphics[width =0.5\linewidth]{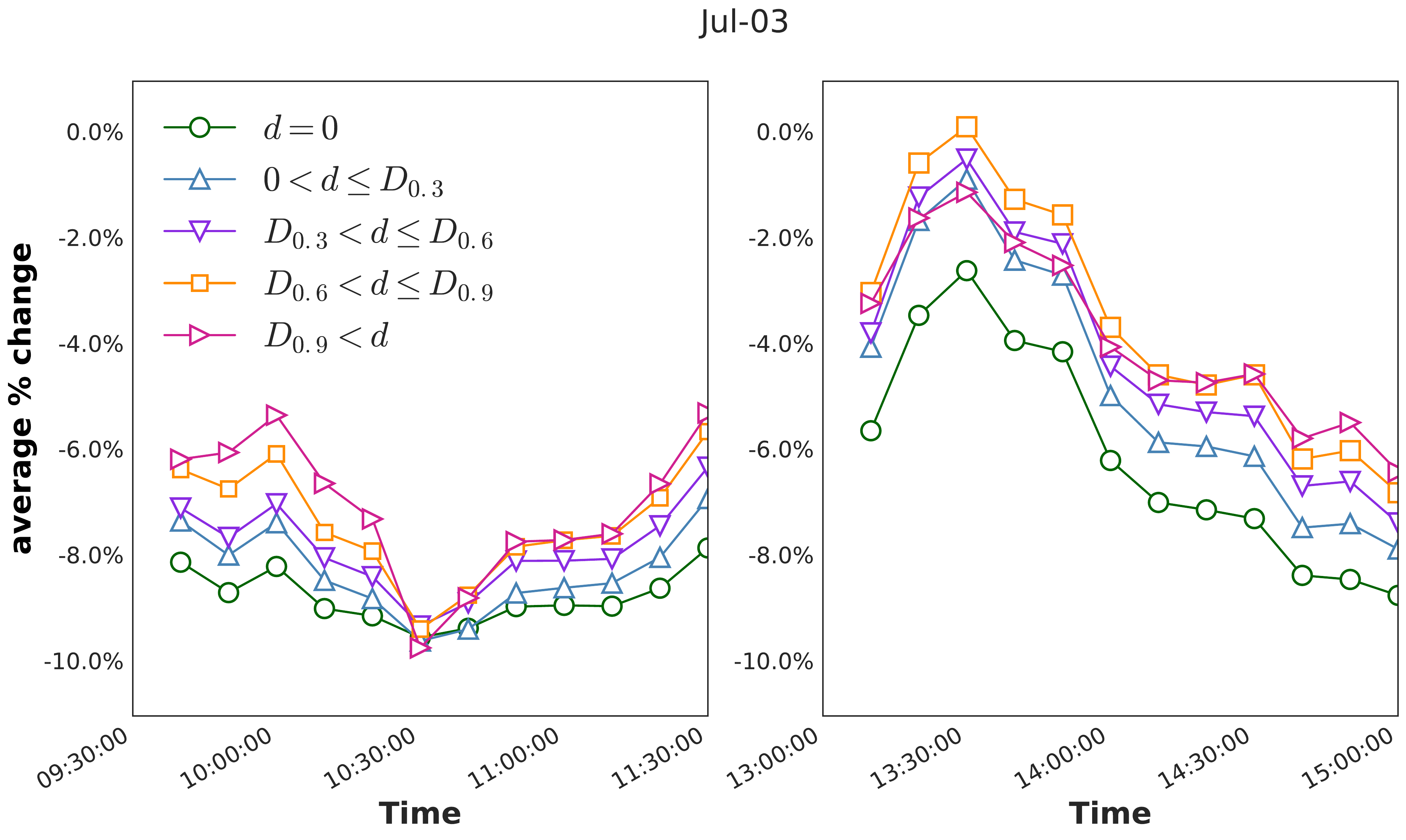}
	\end{minipage}

\caption{\bf{Percentage changes of five groups of stocks on June 29, July 2, and July 3 in 2015.} \normalfont{These three days all experienced sharp market fall. The figures are consistent with Fig.~\ref{fig:626_time} in the main text.}}\label{fig:crisis3days_return}
\end{figure}

\begin{figure}[ht!]
	\begin{minipage}{0.5\linewidth}
		\centering
		\includegraphics[width =0.95\linewidth]{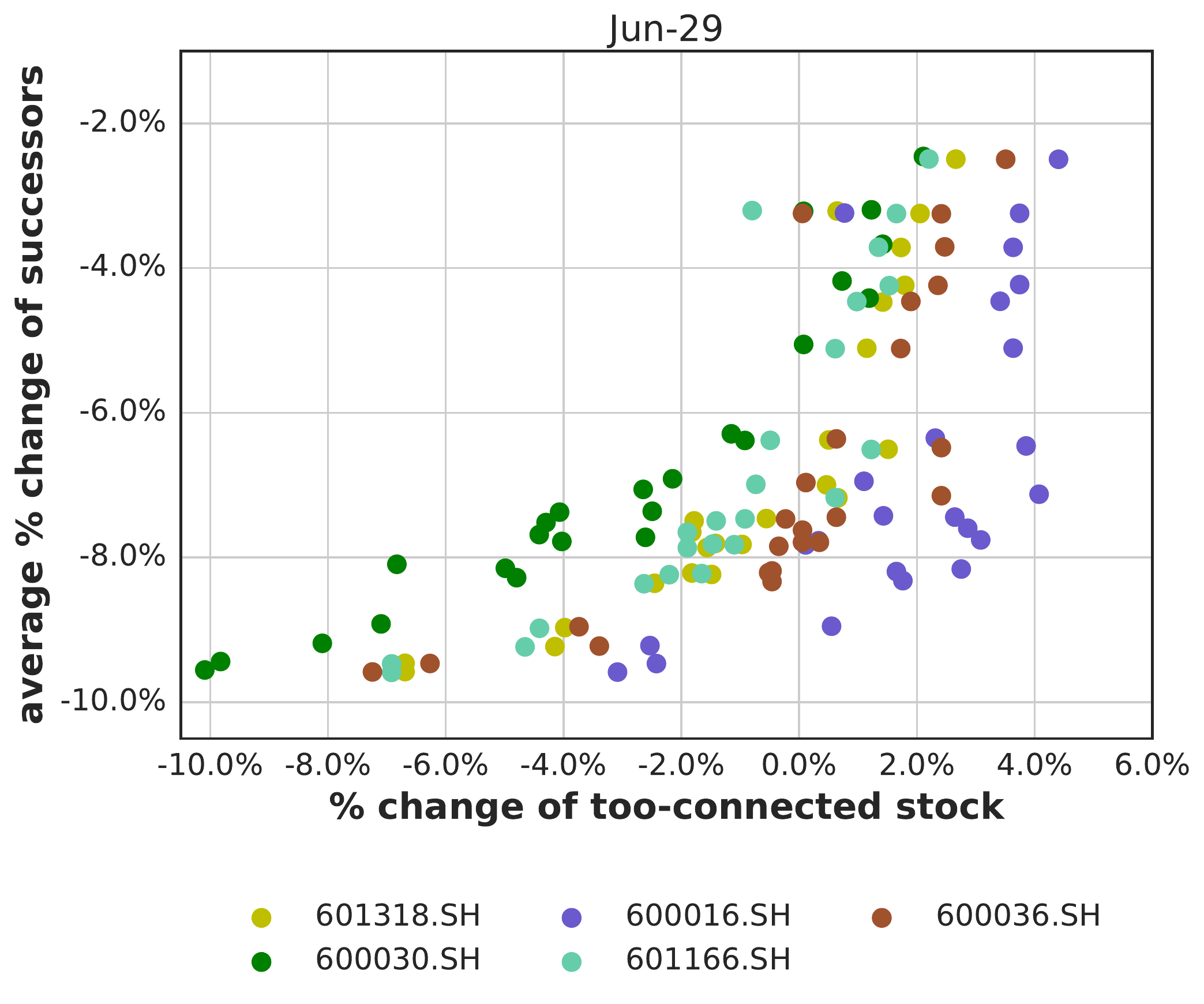}
	\end{minipage}\hfill
	\medskip
	\begin{minipage}{0.5\linewidth}
		\centering
		\includegraphics[width =0.95\linewidth]{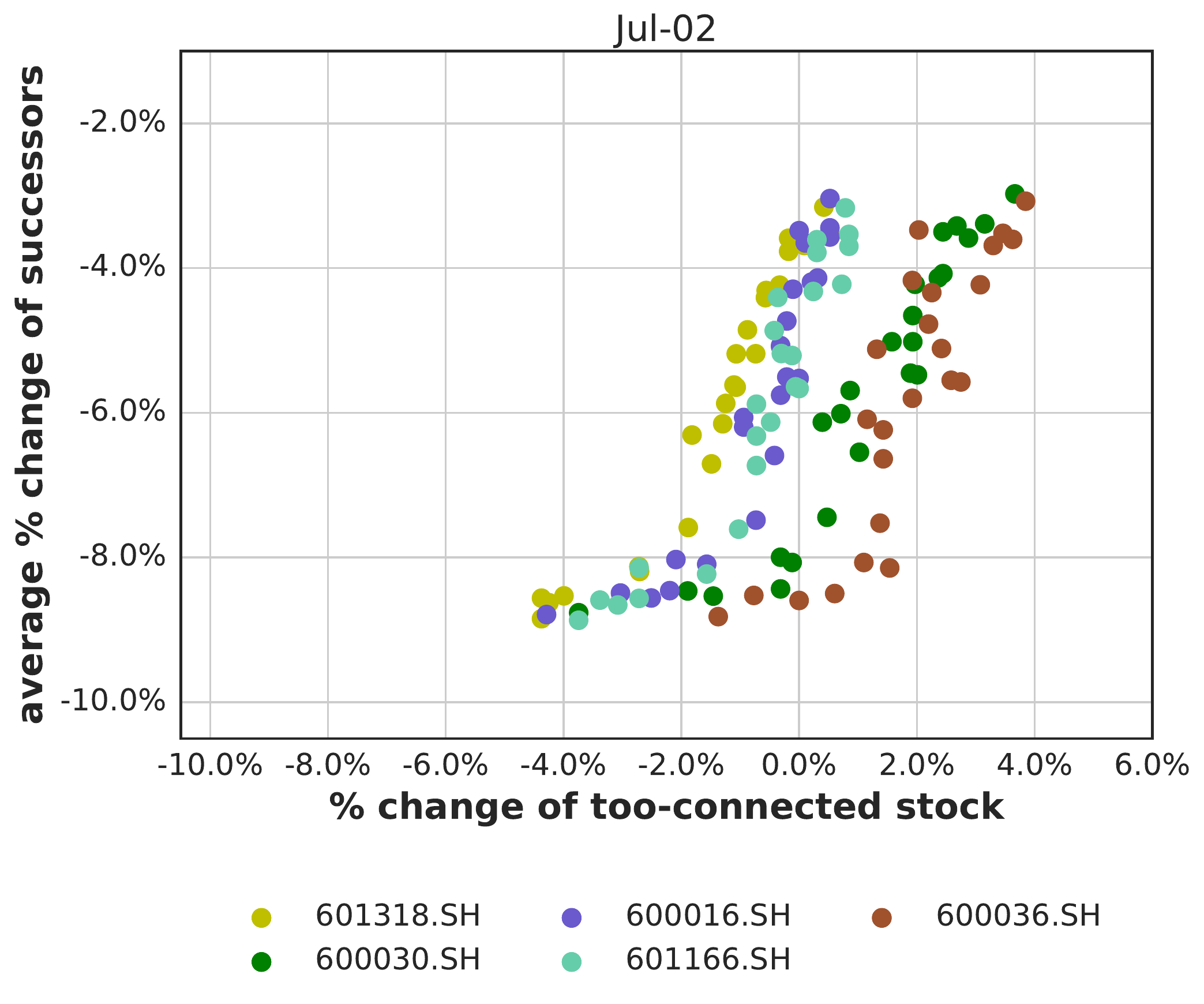}
	\end{minipage}
	\begin{minipage}{\linewidth}
		\centering
		\includegraphics[width =0.5\linewidth]{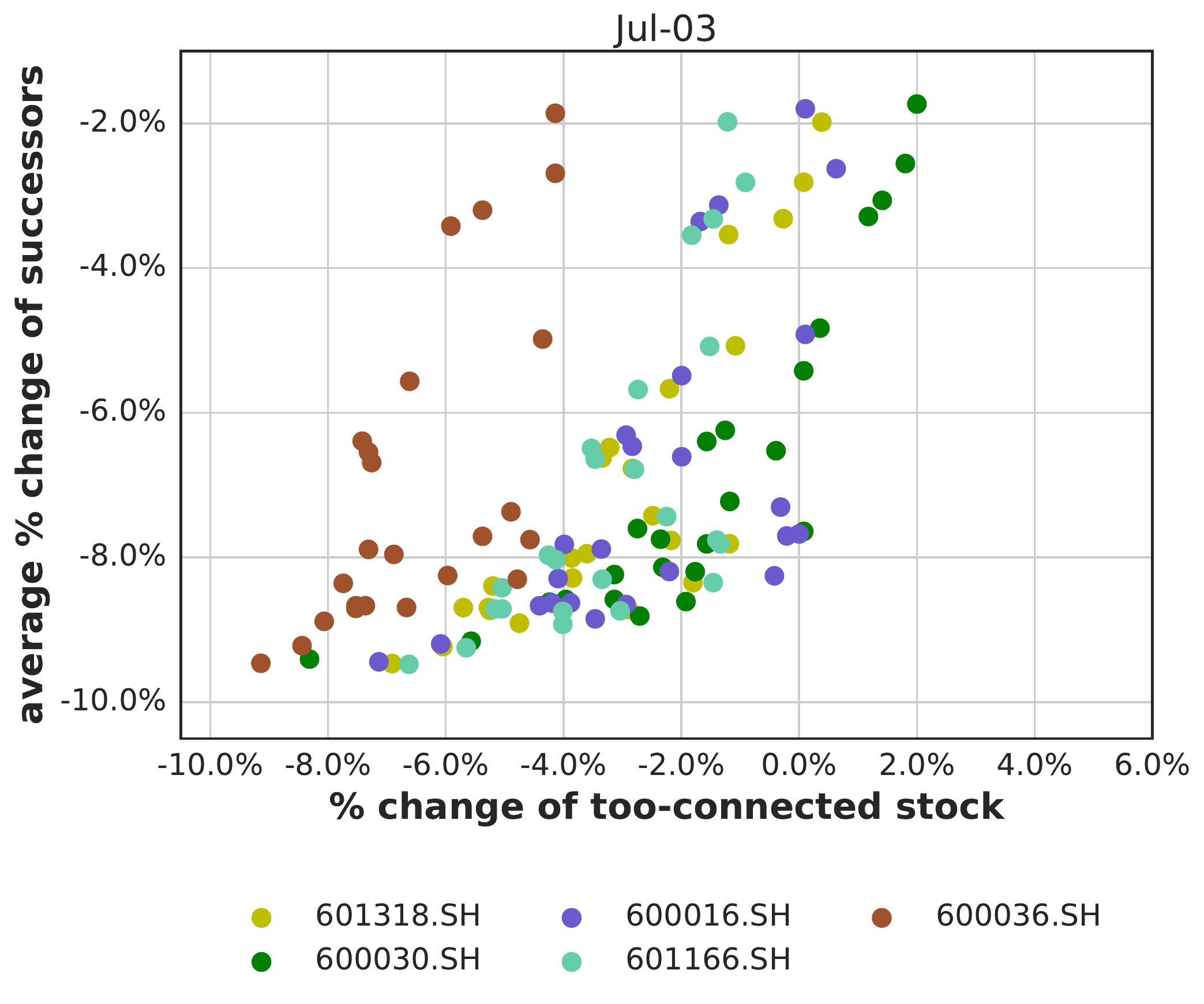}
	\end{minipage}

\caption{\bf{Correlations between `too-connected' stocks' percentage changes and their successors' on June 29, July 2, and July 3 in 2015.} \normalfont{These three days all experienced sharp market falls. The figures are consistent with Fig.~\ref{fig:top5_successor} in the main text.}}\label{fig:crisis3days_scatter}
\end{figure}

\end{appendices}

\end{document}